\documentclass[onecolumn]{aastex}

\usepackage{graphicx}
\usepackage{bm}
\usepackage{color}
\usepackage{amsfonts}
\usepackage{amssymb}
\usepackage{amsmath}

\input epsf

\begin{document}

\title{\Large{Constraining Parameters of Generalized Cosmic Chaplygin Gas in Loop Quantum Cosmology}}

\author{{\bf Chayan Ranjit}\altaffilmark{1}}
\author{{\bf Ujjal Debnath}\altaffilmark{2}}

\altaffiltext{1}{{Department of Mathematics, Seacom Engineering
College, Howrah - 711 302, India. Email: chayanranjit@gmail.com}}

\altaffiltext{2}{{Department of Mathematics, Indian Institute of
Engineering Science and Technology, Shibpur, Howrah-711 103,
India. Email: ujjaldebnath@yahoo.com}}

\date{\today}

\begin{abstract}
We have assumed the FRW universe in loop quantum cosmology (LQC)
model filled with the dark matter and the Generalized Cosmic
Chaplygin gas (GCCG) type dark energy where dark matter follows
the linear equation of state. We present the Hubble parameter in
terms of the observable parameters $\Omega_{m0}$ and $H_{0}$ with
the redshift $z$ and the other parameters like $A$, $B$, $w_{m}$,
$ \omega$ and $\alpha$ which coming from our model. From Stern
data set (12 points)\& SNe Type Ia 292 data (from
\cite{Riess1,Riess2,Astier}) we have obtained the bounds of the
arbitrary parameters by minimizing the $\chi^{2}$ test. The
best-fit values of the parameters are obtained by 66\%, 90\% and
99\% confidence levels. Next due to joint analysis with Stern+BAO
and Stern+BAO+CMB observations, we have also obtained the bounds
of the parameters ($A,B$) by fixing some other parameters
$\alpha$, $w_{m}$ and $\omega$. From the best fit values of the
parameters, we have obtained the distance modulus $\mu(z)$ for our
theoretical GCCG model in LQC and from Supernovae Type Ia (union2
sample 552 data from [\cite{Amanullah}] \& Riess 292 data from
[\cite{Riess1,Riess2,Astier}] ), we have concluded that our model
is in agreement with the Supernovae Type Ia sample data. In
addition, we have investigated in details about the various types
of Future Singularities that may be formed in this model and it is
notable that our model is completely free from any types of future
singularities.
\end{abstract}

\maketitle

\section{\normalsize\bf{Introduction}}
It is known from recent observational study that our universe is
expanding with an acceleration and that supported by different
observations of the SNeIa [\cite{Perlmutter,Perlmutter1,Riess,
Riess1}], baryon acoustic oscillations (BAO) [\cite{Eisenstein}],
large scale redshift surveys [\cite{Bachall,Tedmark}], the
measurements of the cosmic microwave background (CMB)
[\cite{Miller,Bennet}], WMAP [\cite{Briddle,Spergel,Spergel1}] and
effects of weak lensing [\cite{Jain}]. The recent trend among the
researchers is that to find the methodology that triggers late
inflation and for that researchers are mainly divided into two
groups, one considering a modification in the geometry by
adjusting the form of original general theory of relativity and
other invoking any mysterious fluid in the form of an evolving
cosmological constant or a quintessential [\cite{Peebles}] type of
scalar field . Those unknown mysterious fluid which has the
property that the positive energy density and sufficient negative
pressure, known as dark energy (DE) [\cite{Paddy,Sahni}] in which
the potential dominates over the kinetic term. In present time, DE
related problems are most interesting research topic of
theoretical physics [\cite{Weinberg}]. There are several
interesting form of solution of this type problem such as phantom
[\cite{Caldwell,Fu}] tachyon scalar field [\cite{Sen,Balart,
Farajollahi,del}], hessence [\cite{Wei}], dilaton scalar field
[\cite{Morris,Marcus}], K-essence scalar field
[\cite{Armen,Bouhmadi, Malquarti}], DBI essence scalar field
[\cite{Spalinski, Martin}] and many others.

Another unknown missing matter component of the universe is known
as the dark matter (DM) which holds together the galaxy clusters.
DM is also needed to explain the current large scale structure of
the universe. It can be predicted that in cosmic concordance
$\Lambda$CDM model, the Universe is formed of 26\% matter
(baryonic + dark matter) and $\sim$ 74\% of a smooth vacuum energy
component, whereas the thermal CMB component contributes only
about 0.01\%, however, its angular power spectrum of temperature
encode important information about the structure formation process
and other cosmic observables.

Loop Quantum Gravity (LQG) theory has been basically trying to
quantize the gravity with a non-perturbative and background
independent way. As a result, the quantum effect of our universe
quite comfortably describe by LQG [\cite{Ashtekar, Rovelli}]. The
theory and principles of LQG when combined with cosmological
framework then it creates a new theoretical framework, named as
Loop Quantum Cosmology(LQC) [\cite{Bojowald2001, Bojowald2002,
Bojowald2005, Bojowald2008, Ashtekar2007, Ashtekar2003,
Ashtekar2006, Ashtekar2008}]. LQC is basically based on discrete
quantum geometry instead of classical space-time continuum.
Friedmann equation is modified by adding a term quadratic in
density to describe the effect of LQG. In LQC, the standard
Friedmann equation is modified with the help of the
non-perturbative effects which leads to the correction term
$\frac{\rho^{2}}{\rho_{c}}$ and which leads to the result of
mechanically bouncy universe when the matter energy density
reaches to the level of Plank density $\rho_{c}$. In 2005,
Bojowald [\cite{Bojowald2005}] reviewed to give an overview and
summary of the current status of the research work on LQC in
detail and that review was also modified by \cite{Bojowald2008}. A
valuable report about the existing state of art on LQC is
discussed in \cite{Ashtekar2011}. Recently, Sadjadi
[\cite{Sadjadi}] has been discussed about the related study on LQC
like a super acceleration and its possible phase transitions,
i.e., the crossing of the phantom divide line $\omega=-1$.

In observational study, the theoretical models and bounds of the
parameters are tested by the combinations of different
observations astrophysical data repeatedly. The observational
facts are not explained properly by standard big bang cosmology
with perfect fluid. Even though in Einstein's gravity, the
cosmological constant $\Lambda$ (which has the equation of state
$w_{\Lambda}=-1$) allows the cosmic acceleration at late times,
but till now there is no proof of the origin of $\Lambda$ and the
observational bounds on $\Lambda$ are incompatible with
theoretical predictions in vacuum state.

For flat universe, if we assume the universe is filled with
dust-like matter and dark energy, then we need to know
$\Omega_{m}$ of the dust-like matter and $H(z)$ to a very high
accuracy in order to get a handle on $\Omega_{X}$ or $w_{X}$ of
the dark energy [\cite{Paddy1,Paddy2}]. From observations, this
can be a fairly degeneracy for determining $w_{X}(z)$. For
$z>0.01$, TONRY data set with the 230 data points [\cite{Tonry}]
with the 23 points from Barris et al [\cite{Barris}] are still
valid. For $1 < z < 1.6$, the ``gold'' sample of Riess et al
[\cite{Riess1}] with 156 data points are valid. In the flat FRW
universe, one finds $\Omega_{\Lambda}+\Omega_{m}=1$, which are
currently favoured by CMBR data (for recent WMAP results, see
[\cite{Spergel}]). For the most recent Riess data set gives a
best-fit value of $\Omega_{m}$ to be $0.31\pm 0.04$, which matches
with the value $\Omega_{m}=0.29^{+0.05}_{-0.03}$ obtained by Riess
et al [\cite{Riess}]. In comparison, the best-fit $\Omega_{m}$ for
flat models was found to be $0.31\pm 0.08$ [\cite{Paddy1}]. The
best-fit constant equation of state parameter $w$ for Union 2 data
sample gives
$w=-0.997^{+0.050}_{-0.054}$(stat)$^{+0.077}_{-0.082}$(stat+sys
together) for a flat universe, or
$w=-1.038^{+0.056}_{-0.059}$(stat)$^{+0.093}_{-0.097}$(stat+sys
together) with curvature [\cite{Amanullah}]. Now, Chaplygin gas is
the more effective candidate of dark energy with equation of state
$p=-B/\rho$ [\cite{Kamenshchik}] with $B>0$. It has been
generalized to the form $p=-B/\rho^{\alpha}$ [\cite{Gorini}] and
thereafter modified to the form $p=A\rho-B/\rho^{\alpha}$
[\cite{Debnath}]. The MCG best fits with the 3 year WMAP and the
SDSS data with the choice of parameters $A =0.085$ and $\alpha =
1.724$ [\cite{Lu}] which are improved constraints than the
previous ones $-0.35 < A < 0.025$ [\cite{Jun}].

In this work, we assume the FRW universe in Loop Quantum Cosmology
(LQC) model filled with the dark matter and the MCG type dark
energy. We present the Hubble parameter in terms of the observable
parameters $\Omega_{m0}$ and $H_{0}$ with the redshift $z$ and the
other parameters like $A$, $B$, $w_{m}$, $ \omega$ and $\alpha$ in
Section 3. From Stern data set (12 points), we obtain the bounds
of the arbitrary parameters by minimizing the $\chi^{2}$ test in
Subsection 3.1. The best-fit values of the parameters are obtained
by 66\%, 90\% and 99\% confidence levels. Next due to joint
analysis with BAO and CMB observations, we also obtain the bounds
and the best fit values of the parameters ($A,B$) by fixing some
other parameters $H_{0}, \Omega_{m0}, w_{m},\omega$ and $\alpha$
at their most suitable values in Subsection 3.2 and Subsection 3.3
respectively. From the best fit of distance modulus $\mu(z)$ for
our theoretical MCG model in LQC with SNe Type Ia union2 sample
552 data from [\cite{Amanullah}] in Subsection 3.4, we conclude
that our model is in agreement with the union2 sample data. After
that in section 4 we consider the SNe Type Ia Riess 292 data from
[\cite{Riess1, Riess2,Astier}] and examine the bounds of the
arbitrary parameters $A$ \& $B$ by minimizing the $\chi^{2}$ test
for 66\%, 90\% and 99\% confidence levels by fixing $H_{0},
\Omega_{m0}, w_{m},\omega$ and $\alpha$ at their most suitable
values and then we draw the distance modulus $\mu(z)$ for our
theoretical MCG model in LQC with SNe Type Ia Riess 292 data from
[\cite{Riess1, Riess2,Astier}] in Subsection 4.1 and also
concluded that our model is in agreement with the Riess 292 sample
data. The different types of singularities of this scenario have
been studied in Section 5 and finally, the concluding remarks of
the paper are summarized in Section 6.

\section{BASIC EQUATIONS AND SOLUTIONS FOR GCCG IN LQC}
In recent years, loop quantum gravity (LQG) is outstanding effort
to describe the quantum effect of our universe. Nowadays several
dark energy models are studied in the framework of LQC. Till now,
Quintessence and phantom dark energy models [\cite{Wu,Chen}] have
been studied in the cosmological evolution in LQC. Then Modified
Chaplying Gas coupled to dark matter in the universe and it was
described in the frame work LQC by Jamil et al [\cite{jamil}] who
resolved the famous cosmic coincidence problem in modern
cosmology. Some authors have studied the model with an interacting
phantom scalar field with an exponential potential and deduced
that the dark energy dominated future singularities have been
appearing in the standard FRW cosmology but some of these
singularities may be avoided by loop quantum effects.

We consider the flat homogeneous and isotropic universe described
by FRW metric, so the modified Einstein's field equations in LQC
are given by [\cite{jamil}]
\begin{equation}
H^{2}=\frac{\rho}{3}\left(1-\frac{\rho}{\rho_{c}}\right)
\end{equation}
and
\begin{equation}
\dot{H}=-\frac{1}{2}(\rho+p)\left(1-\frac{2\rho}{\rho_{c}}\right)
\end{equation}
 where $H$ is the Hubble parameter defined as
$H=\frac{\dot{a}}{a}$ with $a$ is the scale factor. Where
$\rho_{c}=\sqrt{3}\pi^{2}\gamma^{3}G^{2}\hbar$ is called the
critical loop quantum density, $\gamma$ is the dimensionless
Barbero-Immirzi parameter. Here the universe begins to bounce and
then oscillate forever when the energy density of the universe
becomes of the same order of the critical density $\rho_{c}$. Thus
the big bang, big rip and other singularities problems, which
could not explained by the Einstein's cosmology, might solve in
LQC. It is to be noted that the parameter $\gamma$ is fixed in LQC
by the requirement of the validity of Bekenstein-Hawking entropy
for the Schwarzschild black hole and it has been suggested that
$\gamma \sim 0.2375$ by the black hole thermodynamics in LQC. The
physical solutions are allowed only when $\rho\le\rho_{c}$. For
$\rho=\rho_{c}$, it is called bounce. The maximum value of the
Hubble factor $H$ is settled for $\rho_{max}=\frac{\rho_{c}}{2}$
and the maximum value of Hubble factor is
$\frac{\kappa\rho_{c}}{12}$.\\Here $\rho=\rho_{x}+\rho_{m}$ and
$p=p_{x}+p_{m}$, where $\rho_{m}$, $p_{m}$ are the matter-density
and pressure contribution of matter respectively and $\rho_{x}$,
$p_{x}$ are respectively the energy density and pressure
contribution of some dark energy. Now we consider the Universe is
filled with Generalized Cosmic Chaplygin Gas (GCCG) model whose
equation of state (EOS) is given by [\cite{Gon,Writ}]
\begin{equation}
p_{x} = -\rho_{x}^{-\alpha}[C+(\rho_{x}^{1+\alpha}-C)^{-\omega}]
\end{equation}
where $C=\frac{A}{1+\omega}-1$ with $A$ is a constant which can
take on both positive and negative values and $-l<\omega<0$, $l$
being a positive definite constant which can take on values larger
than unity. We also consider the dark matter and the dark energy
are separately conserved and the conservation equations of dark
matter and dark energy (GCCG) are given by
\begin{equation}
\dot{\rho}_{m}+3H(\rho_{m}+p_{m})=0
\end{equation}
and
\begin{equation}
    \dot{\rho}_{x}+3H(\rho_{x}+p_{x})=0
\end{equation}
From first conservation equation (4) we have the solution of
$\rho_{m}$ as
\begin{equation}
\rho_{m}=\rho_{m0}(1+z)^{3(1+w_{m})}
\end{equation}
where $p_{m}=\rho_{m}w_{m}$. From the conservation equation (5) we
have the solution of the energy density as
\begin{equation}
\rho_{x}=\left[\left(\frac{A}{1+\omega}-1\right)+\left(1+B(1+z)^{3(1+\alpha)(1+\omega)}\right)^{\frac{1}{1+\omega}}\right]^{\frac{1}{1+\alpha}}
\end{equation}
where $B$ is the integrating constant, $z=\frac{1}{a}-1$ is the
cosmological redshift (choosing $a_{0}=1$) and the first constant
term can be interpreted as the contribution of dark energy.

\section{\bf{Observational Data Analysis}}
From the solution (7) of GCCG and defining the dimensionless
density parameter $\Omega_{m0}=\frac{\rho_{m0}}{3 H_{0}^{2}}$, we
have the expression for Hubble parameter $H$ in terms of redshift
parameter $z$ as follows ($8\pi G=c=1$)
\begin{eqnarray*}
H(z)=\frac{1}{\sqrt{3}}\left[3\Omega_{m0}H_{0}^{2}(1+z)^{3(1+w_{m})}+\left(\left(\frac{A}{1+\omega}-1\right)
\left(1+B(1+z)^{3(1+\alpha)(1+\omega)}\right)^{\frac{1}{1+\omega}}\right)\right]^{\frac{1}{2}}
\end{eqnarray*}
\begin{equation}
\times\left[1-\frac{3\Omega_{m0}H_{0}^{2}(1+z)^{3(1+w_{m})}+\left(\left(\frac{A}{1+\omega}-1\right)
\left(1+B(1+z)^{3(1+\alpha)(1+\omega)}\right)^{\frac{1}{1+\omega}}\right)}{\rho_{c}}\right]^{\frac{1}{2}}
\end{equation}
From equation (8), we see that the value of $H$ depends on
$H_{0},A,B,\Omega_{m0},w_{m},\omega,\alpha,z$. The $E(z)$ can be
written as
\begin{equation}
E(z)=\frac{H(z)}{H_{0}}
\end{equation}
Now $E(z)$ contains unknown parameters like $A,B,\omega$ and
$\alpha$. Now we will fixing two parameters and by observational
data set the relation between the other two parameters will obtain
and find the bounds of the parameters.
\[
\begin{tabular}{|c|c|c|}
\hline
  ~~~~~~$z$ ~~~~& ~~~~$H(z)$ ~~~~~& ~~~~$\sigma(z)$~~~~\\
  \hline
  0 & 73 & $\pm$ 8 \\
  0.1 & 69 & $\pm$ 12 \\
  0.17 & 83 & $\pm$ 8 \\
  0.27 & 77 & $\pm$ 14 \\
  0.4 & 95 & $\pm$ 17.4\\
  0.48& 90 & $\pm$ 60 \\
  0.88 & 97 & $\pm$ 40.4 \\
  0.9 & 117 & $\pm$ 23 \\
  1.3 & 168 & $\pm$ 17.4\\
  1.43 & 177 & $\pm$ 18.2 \\
  1.53 & 140 & $\pm$ 14\\
  1.75 & 202 & $\pm$ 40.4 \\ \hline
\end{tabular}
\]
{\bf Table 1:} The Hubble parameter $H(z)$ and the standard error
$\sigma(z)$ for different values of redshift $z$.

In the following subsections, we shall investigate the data
analysis mechanism for Stern, Stern+BAO and Stern+BAO+CMB
observational data to find some bound of the parameters of GCCG
with LQC. We shall use the $\chi^{2}$ minimization technique
(statistical data analysis) to test the theoretical Hubble
parameter with the observed data set to get the best fit values of
the unknown parameters with different confidence levels.

\subsection{Analysis with Stern ($H(z)$-$z$) Data Set}
In 2010, Stern et al [\cite{Stern}] proposed an observed data set
which is known as Stern ($H(z)$-$z$) data set. Stern data set
consisted with the observed value of Hubble parameter $H(z)$ and
the standard error $\sigma(z)$ for different values of redshift
$z$ (twelve data points), which are given in Table 1. Here we use
Stern data set (twelve data points) to analyze the model. Before
going to apply $\chi^{2}$ minimization technique, we first form
the $\chi^{2}$ statistics as a sum of standard normal distribution
as follows:
\begin{equation}
{\chi}_{Stern}^{2}=\sum\frac{(H(z)-H_{obs}(z))^{2}}{\sigma^{2}(z)}
\end{equation}
where $H_{obs}(z)$ and $H(z)$ are observational and theoretical
values of Hubble parameter at different redshifts $z$ respectively
and $\sigma(z)$ is the corresponding error for the particular
observation given in Table 1. Also, the nuisance parameter
$H_{obs}$ can be safely marginalized. Here the present value of
Hubble parameter $H_{0}$ is been settled at 72 $\pm$ 8 Km s$^{-1}$
Mpc$^{-1}$ with a fixed prior distribution. Now we shall determine
the bounds of parameters $A$ and $B$ for different $\alpha$ from
minimizing the above distribution ${\chi}_{Stern}^{2}$. Fixing the
other parameters $\Omega_{m0},w_{m},\omega,\alpha$, the relation
between $A$ and $B$ can be determined by the observational data.
The probability distribution function in terms of the parameters
$A,B,\Omega_{m0},w_{m},\omega$ and $\alpha$ can be written as
\begin{equation}
L= \int e^{-\frac{1}{2}{\chi}_{Stern}^{2}}P(H_{0})dH_{0}
\end{equation}
where $P(H_{0})$ is the prior distribution function for $H_{0}$.

Now, using $\chi^{2}$ minimization technique, we plot the graph of
the unknown parameters $A$ and $B$ for different $\alpha$ and
fixing the other parameters for different confidence levels (like
66\%, 90\% and 99\%). The best fit values of the parameters $A$
and $B$ are written in Table 2. It is to be noted that our best
fit analysis with Stern observational data support the theoretical
range of the parameters.

The 66\% (solid, blue), 90\% (dashed, red) and 99\% (dashed,
black) contours for $(A,B)$ are plotted in figures 1, 2 and 3 for
different values of $\alpha$. Also the best fit values of $A$ and
$B$ are tabulated in Table 2.
\[
\begin{tabular}{|c|c|c|c|}
\hline
 ~~~~~~$\alpha$ ~~~~&~~~~$A$ ~~~~~& ~~~~~~~$B$ ~~~~~~~~& ~~~$\chi^{2}_{min}$~~~~~~\\
  \hline
 0.0020 &  0.628976 & 5.62894 & 7.09652 \\
 0.0010 &  0.628989 & 5.62894 & 7.09652 \\
 0.0005 & -0.069897 & 5.58487 & 7.09670 \\
  \hline
\end{tabular}
\]
{\bf Table 2:} $H(z)$-$z$ (Stern): The best fit values of $A$, $B$
and the minimum values of $\chi^{2}$ for different values of
$\alpha$ and fixed value of other parameters.
\begin{figure}
\includegraphics[height=2in]{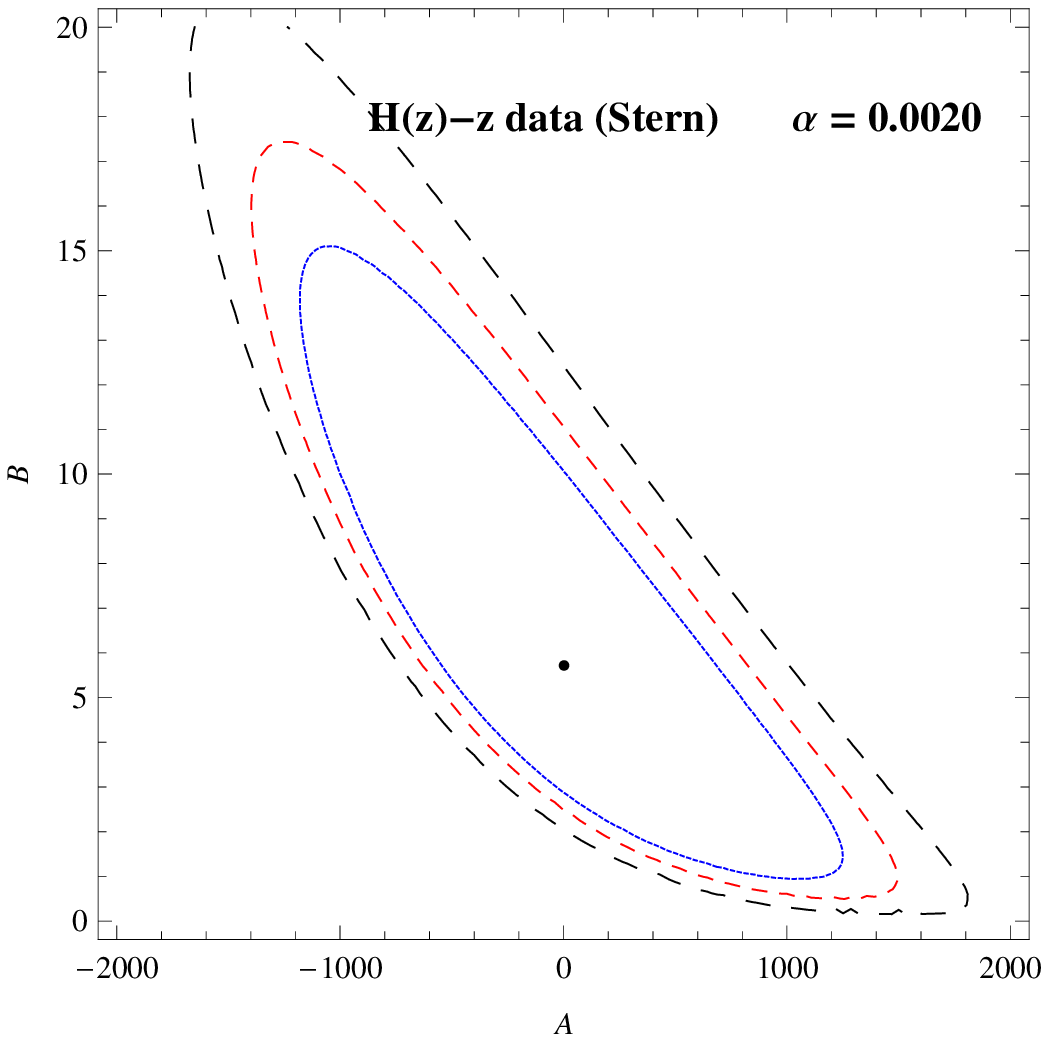}~~
\includegraphics[height=2in]{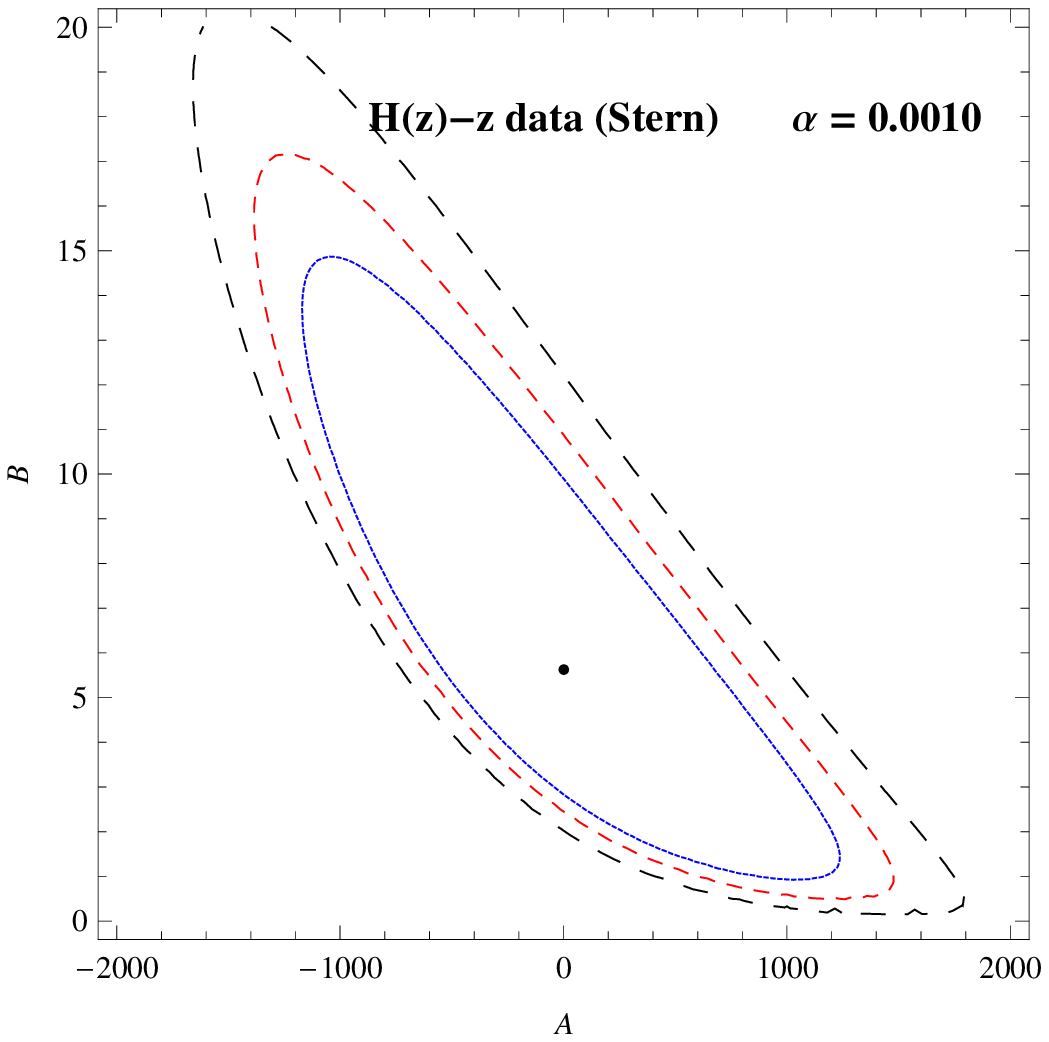}~~
\includegraphics[height=2in]{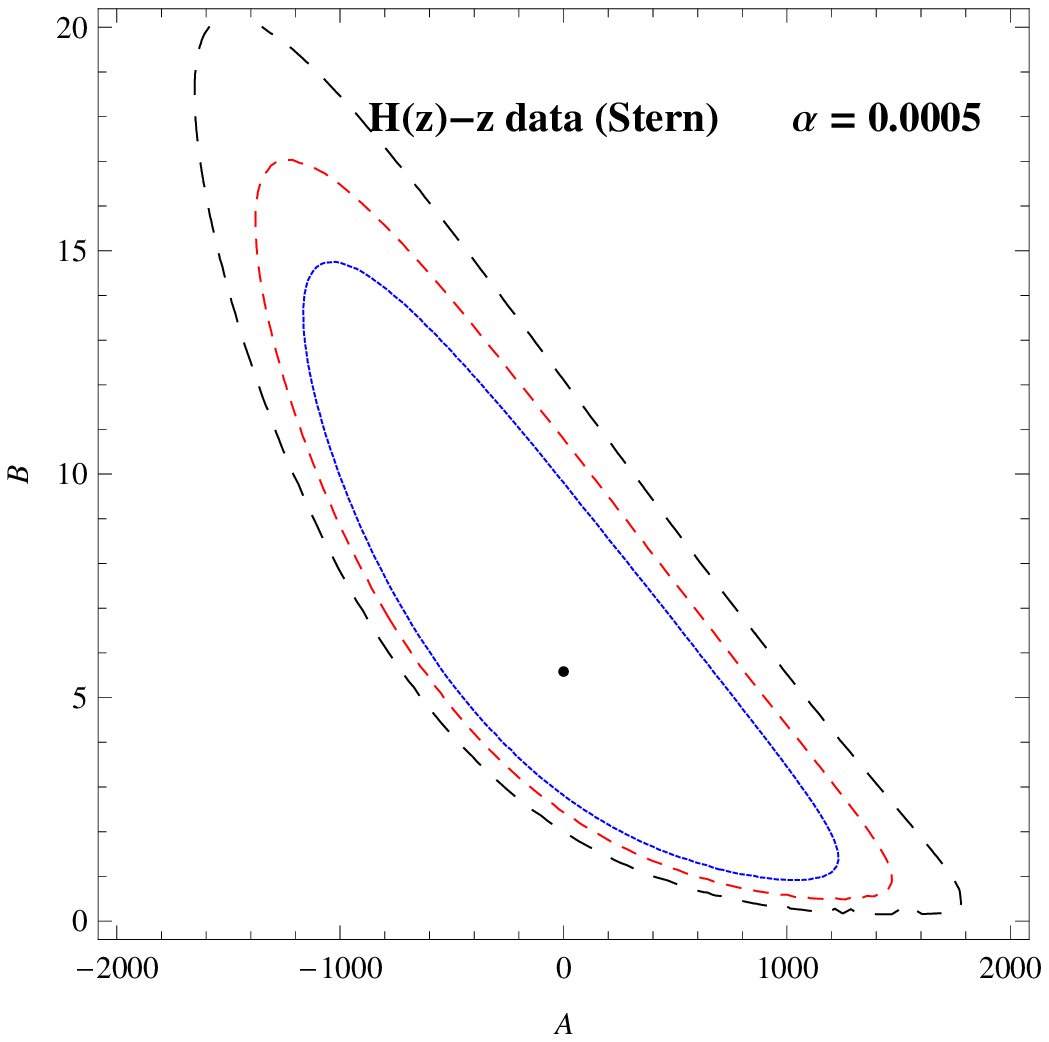}\\
\vspace{2mm}
~~~~~~~~~~~~~~~Fig.1~~~~~~~~~~~~~~~~~~~~~~~~~~~~~~~~~Fig.2~~~~~~~~~~~~~~~~~~~~~~~~~~~~~~~~~~Fig.3~~~~~~\\
\vspace{1mm} Fig.1-3 show that the variation of $A$ with $B$ for
$\Omega_{m0}=0.0643, w_{m}=0.051, \omega=-0.92$ with $\alpha =$
$0.0020$, $0.0010$ \& $0.0005$ respectively for different
confidence levels. The 66\% (solid, blue), 90\% (dashed, red) and
99\% (dashed, black) contours are plotted in these figures for the
$H(z)$-$z$ (Stern) analysis. \vspace{1mm}
\end{figure}

\subsection{Joint Analysis with Stern $+$ BAO Data Sets}
Now we use the statistical approach of joint analysis put
forwarded by Eisenstein et al \cite{Eisenstein}. The Baryon
Acoustic Oscillation (BAO) peak parameter value has been proposed
in their method of joint analysis. Sloan Digital Sky Survey (SDSS)
survey is one of the primordial redshift survey by which the BAO
signal has been directly detected at a scale $\sim$ 100 MPc. In
this case, the said analysis is actually the combination of
angular diameter distance and Hubble parameter at that redshift.
This analysis is independent of the measurement of $H_{0}$ and not
containing any particular dark energy. Here we shall check the
parameters $A$ and $B$ with the measurements of the BAO peak at
low redshift (with range $0<z<0.35$) using standard $\chi^{2}$
technique. The error, corresponding to the standard deviation, is
follow the Gaussian distribution. Low-redshift distance have the
ability to measure the Hubble constant $H_{0}$ directly. It
lightly depends on different cosmological parameters and the
equation of state of dark energy. The BAO peak parameter might be
defined as
\begin{equation}
{\cal
A}=\frac{\sqrt{\Omega_{m}}}{E(z_{1})^{1/3}}\left(\frac{1}{z_{1}}~\int_{0}^{z_{1}}
\frac{dz}{E(z)}\right)^{2/3}
\end{equation}
 Here $E(z)=H(z)/H_{0}$ is the normalized Hubble parameter. The
redshift $z_{1}$ is the typical redshift of the SDSS sample whose
value is settled as $0.35$ and the integration term is the
dimensionless comoving distance to the redshift $z_{1}$. The value
of the parameter ${\cal A}$ for the flat model of the universe is
proposed as \cite{Eisenstein} ${\cal A}=0.469\pm 0.017$ using SDSS
data from luminous red galaxies survey. Now the $\chi^{2}$
function for the BAO measurement can be written as
\begin{equation}
\chi^{2}_{BAO}=\frac{({\cal A}-0.469)^{2}}{(0.017)^{2}}
\end{equation}
Now the total joint data analysis (Stern+BAO) for the $\chi^{2}$
function is defined by \cite{Wu1, Paul, Paul1, Paul2, Paul3, Chak}
\begin{equation}
\chi^{2}_{total}=\chi^{2}_{Stern}+\chi^{2}_{BAO}
\end{equation}
According to our analysis the joint scheme gives the best fit
values of $A$ and $B$ for different $\alpha$ in Table 3. Finally
we draw the contours $A$ vs $B$ for the 66\% (solid, blue), 90\%
(dashed, red) and 99\% (dashed, black) confidence limits depicted
in figures $4-6$ for different values of $\alpha$.
\[
\begin{tabular}{|c|c|c|c|}
\hline
 ~~~~~~$\alpha$ ~~~~&~~~~$A$ ~~~~~& ~~~~~~~$B$ ~~~~~~~~& ~~~$\chi^{2}_{min}$~~~~~~\\
  \hline
  0.0020 & 1.4401851 & 5.71536 & 768.073 \\
  0.0010 & 0.0296015 & 5.62625 & 768.073 \\
  0.0005 & -0.666052 & 5.58219 & 768.074 \\
   \hline
\end{tabular}
\]
{\bf Table 3:} $H(z)$-$z$ (Stern) + BAO : The best fit values of
$A$, $B$ and the minimum values of $\chi^{2}$ for different values
of $\alpha$ and fixed value of other parameters.
\begin{figure}
\includegraphics[height=2in]{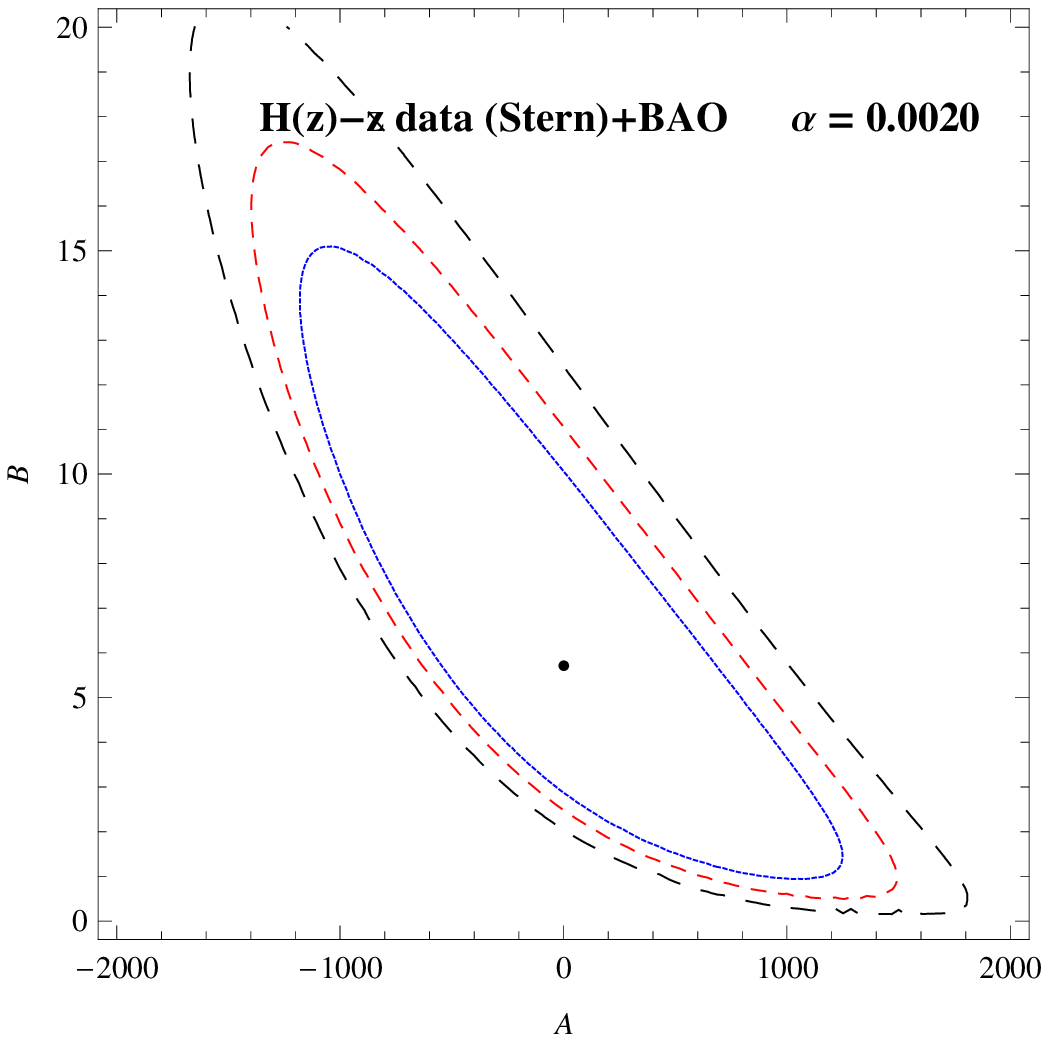}~~
\includegraphics[height=2in]{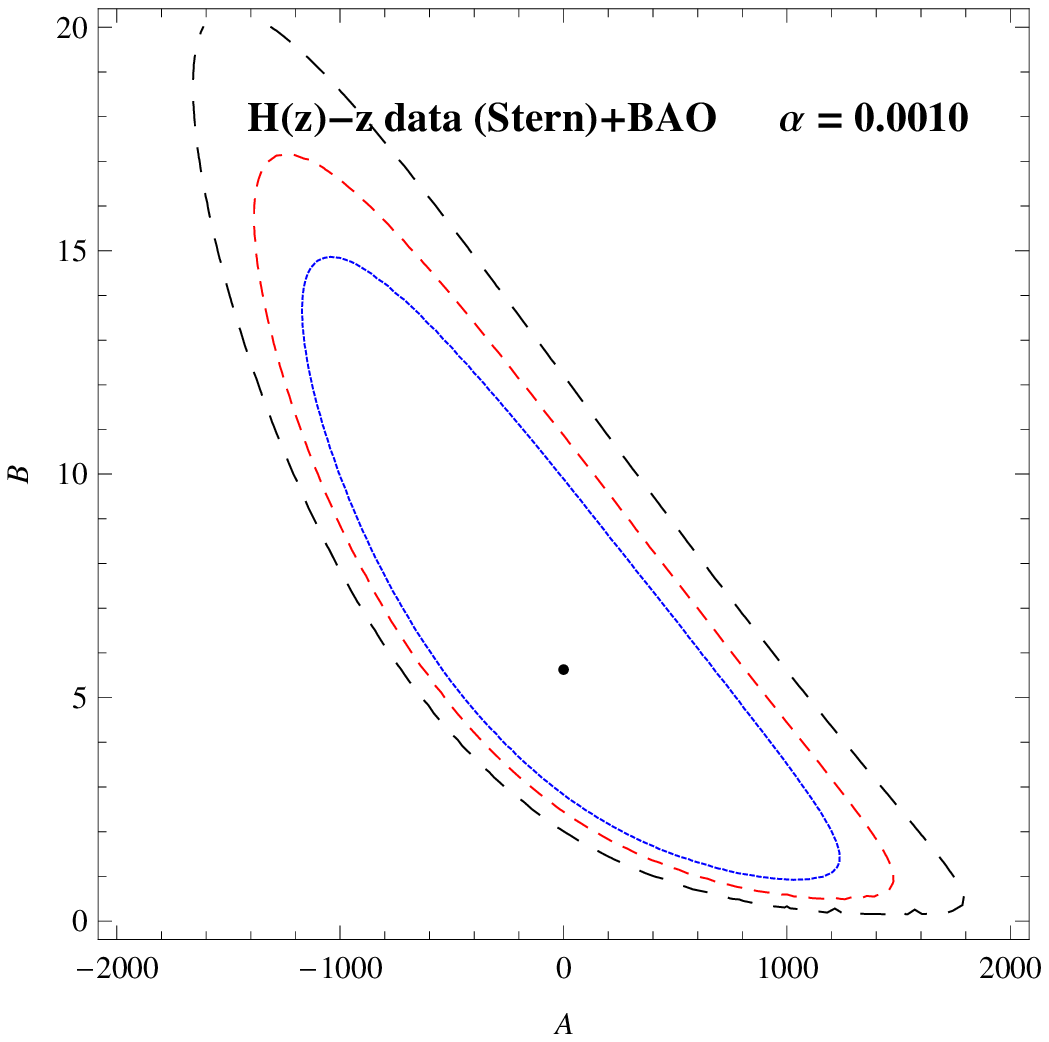}~~
\includegraphics[height=2in]{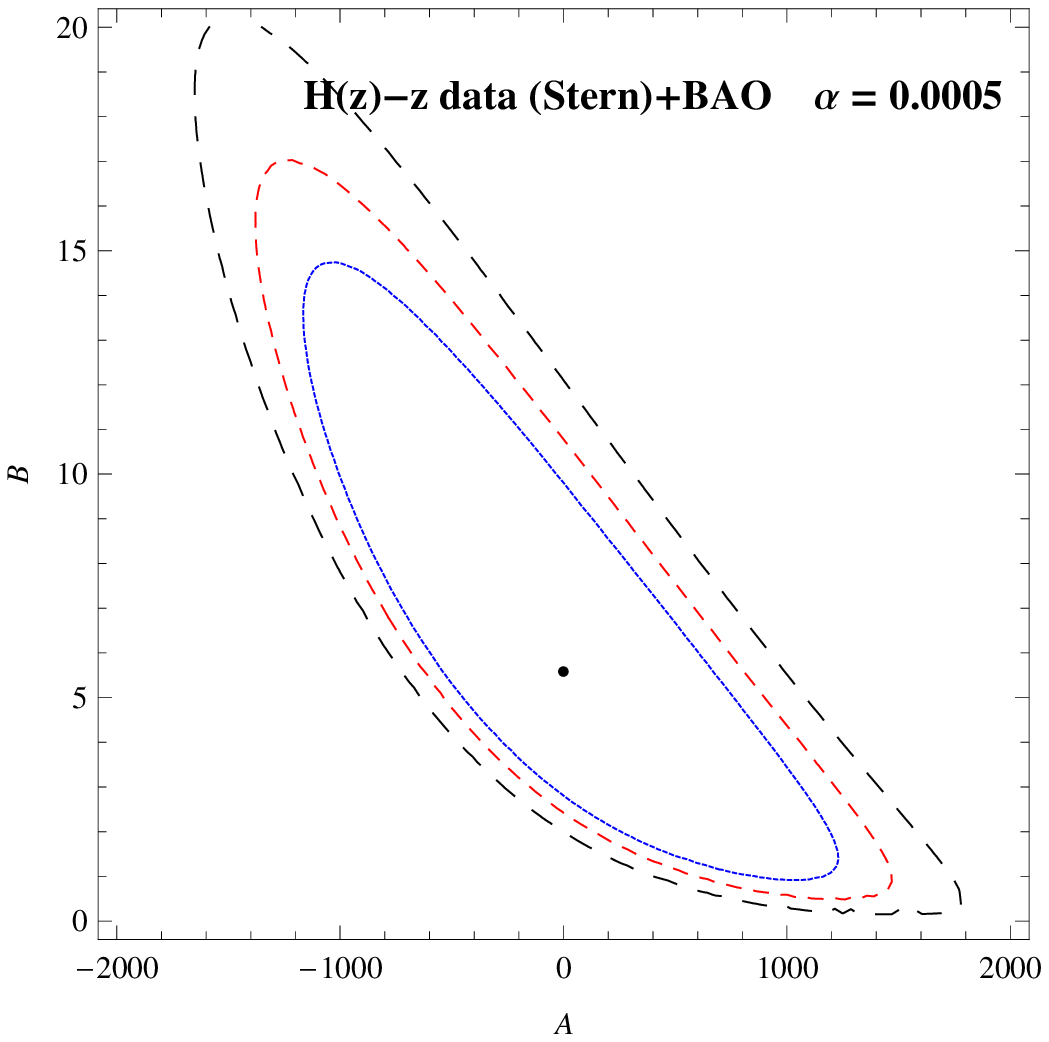}\\
\vspace{1mm}
~~~~~~~~~~~~~~~Fig.4~~~~~~~~~~~~~~~~~~~~~~~~~~~~~~~~~Fig.5~~~~~~~~~~~~~~~~~~~~~~~~~~~~~~~~~~Fig.6~~~~~~\\
\vspace{1mm} Fig.4-6 show that the variation of $A$ with $B$ for
$\Omega_{m0}=0.01, w_{m}=0.051, \omega=-0.92$ with $\alpha =$
$0.0020$, $0.0010$ \& $0.0005$ respectively for different
confidence levels. The 66\% (solid, blue), 90\% (dashed, red) and
99\% (dashed, black) contours are plotted in these figures for the
$H(z)$-$z$(Stern)+BAO joint analysis. \vspace{1mm}
\end{figure}

\subsection{Joint Analysis with Stern $+$ BAO $+$ CMB Data Sets}
In this subsection, we shall follow the pathway, proposed by some
author [\cite{Bond,Efstathiou,Nessaeris}], using Cosmic Microwave
Background (CMB) shift parameter. The interesting geometrical
probe of dark energy can be determined by the angular scale of the
first acoustic peak through angular scale of the sound horizon at
the surface of last scattering which is encoded in the CMB power
spectrum. It is not sensitive with respect to perturbations but
are suitable to constrain model parameter. The CMB power spectrum
first peak is the shift parameter which is given by
\begin{equation}
{\cal R}=\sqrt{\Omega_{m}} \int_{0}^{z_{2}} \frac{dz}{E(z)}
\end{equation}
where $z_{2}$ is the value of redshift at the last scattering
surface.\\ From WMAP7 data of the work of Komatsu et al
\cite{Komatsu} the value of the parameter has proposed as ${\cal
R}=1.726\pm 0.018$ at the redshift $z=1091.3$. Therefore the
$\chi^{2}$ function for the CMB measurement can be written as
\begin{equation}
\chi^{2}_{CMB}=\frac{({\cal R}-1.726)^{2}}{(0.018)^{2}}
\end{equation}
Now when we consider three cosmological tests together, the total
joint data analysis (Stern+BAO+CMB) for the $\chi^{2}$ function
may be defined by
\begin{equation}
\chi^{2}_{TOTAL}=\chi^{2}_{Stern}+\chi^{2}_{BAO}+\chi^{2}_{CMB}
\end{equation}
\begin{figure}
\includegraphics[height=2in]{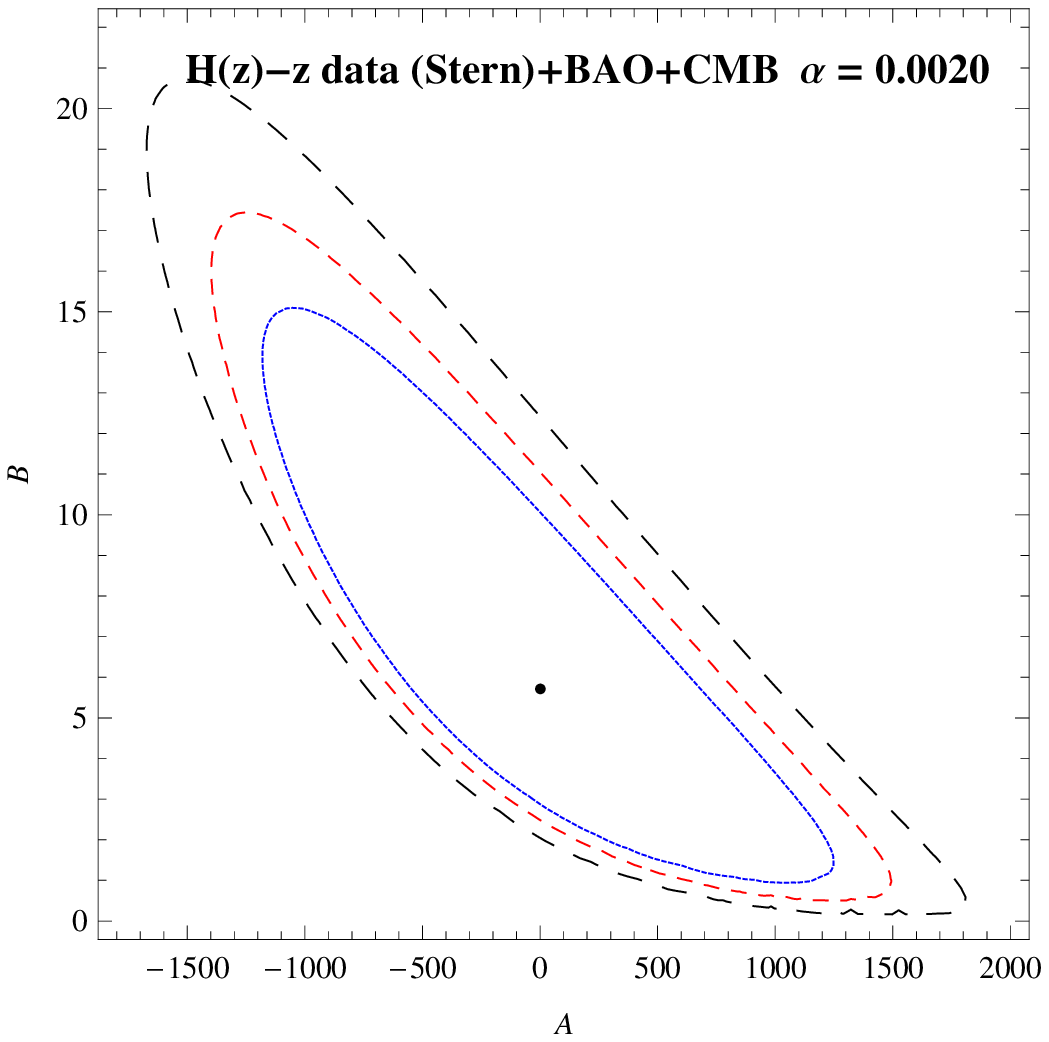}~~
\includegraphics[height=2in]{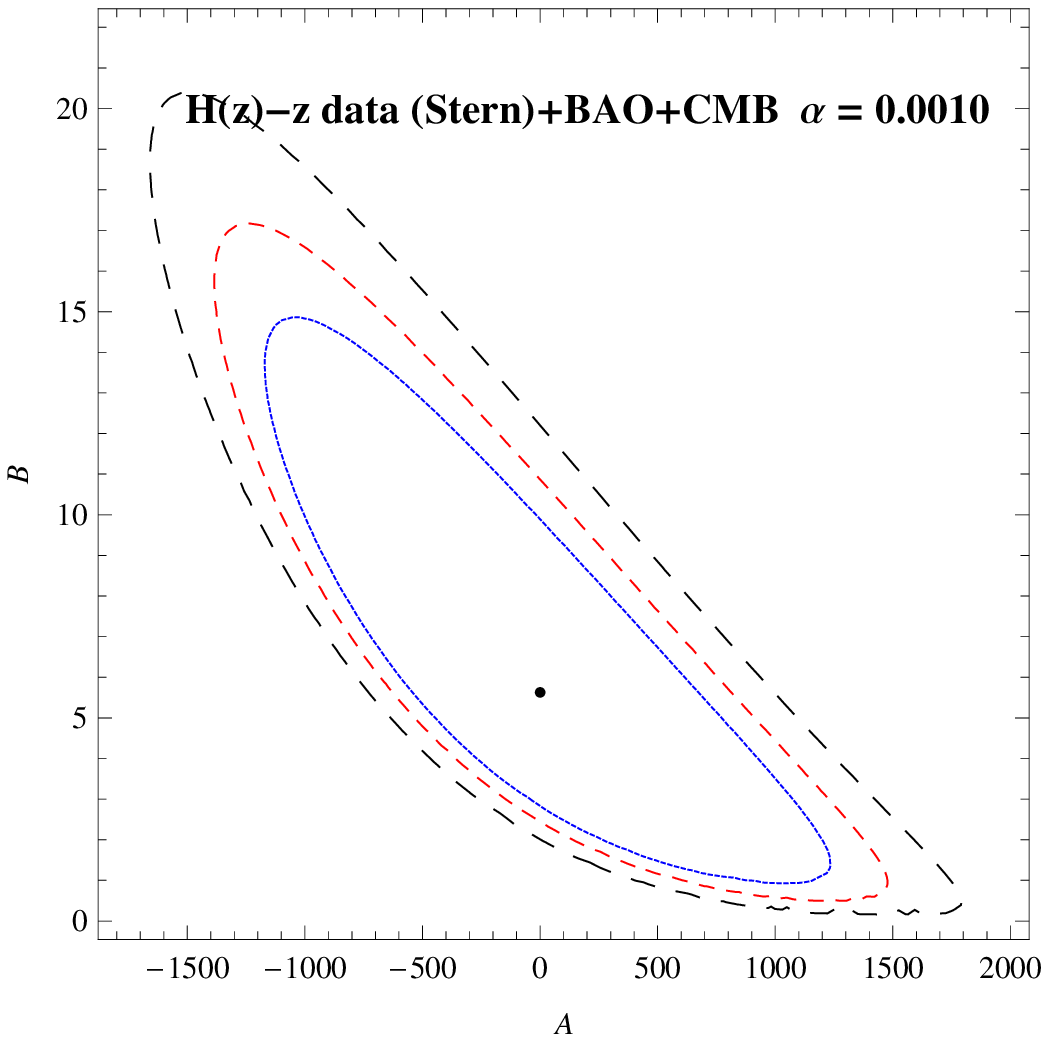}~~
\includegraphics[height=2in]{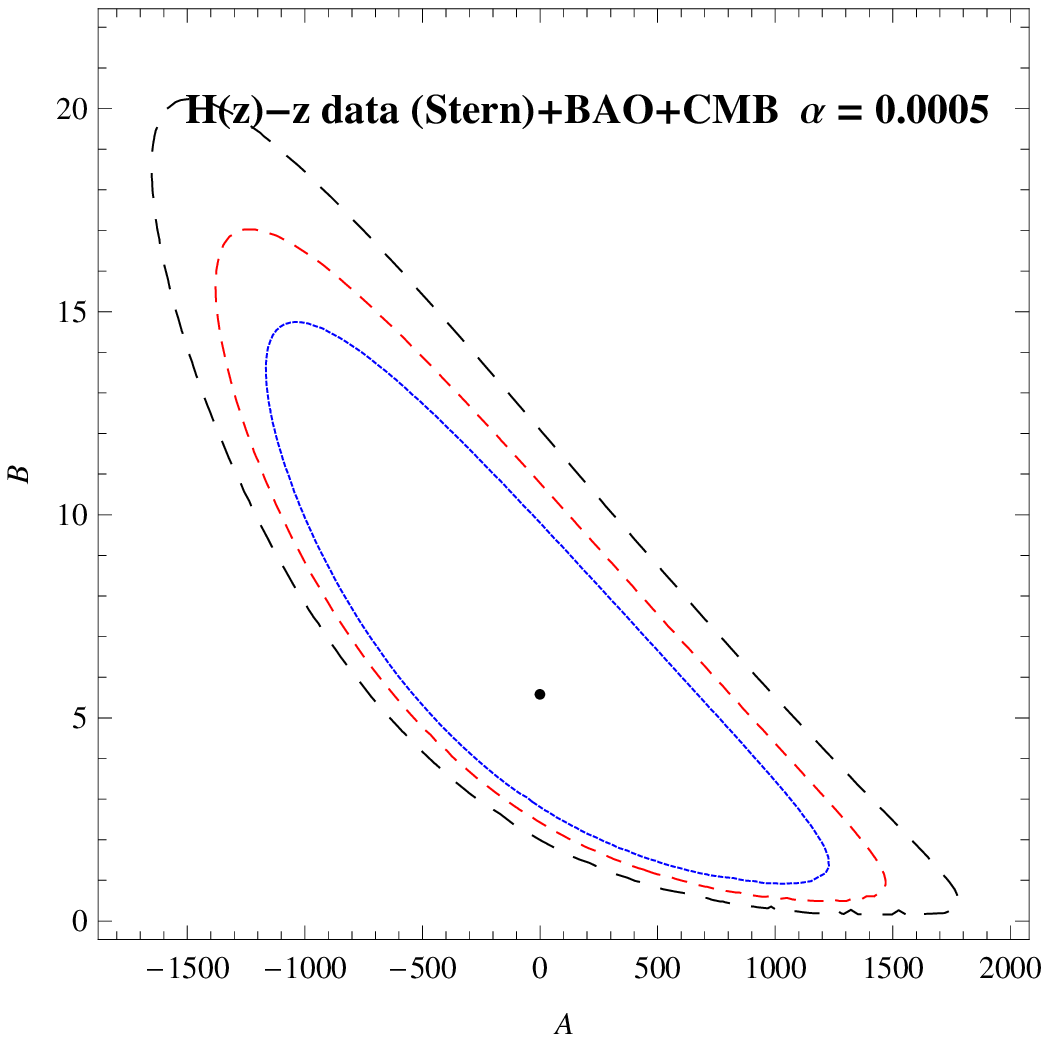}\\
\vspace{1mm}
~~~~~~~~~~~~~~~Fig.7~~~~~~~~~~~~~~~~~~~~~~~~~~~~~~~~~Fig.8~~~~~~~~~~~~~~~~~~~~~~~~~~~~~~~~~~Fig.9~~~~~~\\
\vspace{2mm} Fig.7-9 show that the variation of $A$ with $B$ for
$\Omega_{m0}=0.01, w_{m}=0.051, \omega=-0.92$ with $\alpha =$
$0.0020$, $0.0010$ \& $0.0005$ respectively for different
confidence levels. The 66\% (solid, blue), 90\% (dashed, red) and
99\% (dashed, black) contours are plotted in these figures for the
$H(z)$-$z$(Stern)+BAO+CMB analysis. \vspace{1mm}
\end{figure}
Now the best fit values of $A$ and $B$ for joint analysis of BAO
and CMB with Stern observational data support the theoretical
range of the parameters given in Table 4. The 66\% (solid, blue),
90\% (dashed, red) and 99\% (dashed, black) contours are plotted
in figures 7-9 for different values of $\alpha$.
\[
\begin{tabular}{|c|c|c|c|}
\hline
 ~~~~~~$\alpha$ ~~~~&~~~~$A$ ~~~~~& ~~~~~~~$B$ ~~~~~~~~& ~~~$\chi^{2}_{min}$~~~~~~\\
  \hline
  0.0020 & 1.4816294 & 5.71517 & 9962.75 \\
  0.0010 & 0.0713549 & 5.62606 & 9962.75 \\
  0.0005 & -0.624274 & 5.58201 & 9962.75 \\
  \hline
\end{tabular}
\]
{\bf Table 4:} $H(z)$-$z$ (Stern) + BAO + CMB : The best fit
values of $A$, $B$ and the minimum values of $\chi^{2}$ for
different values of $\alpha$ and fixed value of other parameters.

\subsection{Redshift-Magnitude Observations of Supernovae Type Ia
Union2 552 Sample [From \cite{Amanullah}]}
The main evidence for the existence of dark energy is provided by
the Supernova Type Ia experiments. Two teams of High-$z$ Supernova
Search and the Supernova Cosmology Project have discovered several
type Ia supernovas at the high redshifts [\cite{Perlmutter,
Perlmutter1,Riess,Riess1}] since 1995. The observations directly
measure the distance modulus of a Supernovae and its redshift $z$
[\cite{Riess2,Kowalaski}]. Here we take recent observational data,
including SNe Ia which consists of 557 data points and belongs to
the Union2 sample [\cite{Amanullah}].

Motivated by the work of some authors [\cite{Paul, Paul1, Paul2,
Paul3, Chak}] here we determine distance modulus $d_{L}(z)$ for
our theoretical GCCG in LQC model and tested with the SNe Type Ia
data. From the observations, the luminosity distance $d_{L}(z)$
determines the dark energy density and is defined by
\begin{equation}
d_{L}(z)=(1+z)H_{0}\int_{0}^{z}\frac{dz'}{H(z')}
\end{equation}
and the distance modulus (distance between absolute and apparent
luminosity of a distance object) for Supernovas is given by
\begin{equation}
\mu(z)=5\log_{10} \left[\frac{d_{L}(z)/H_{0}}{1~
\text{MPc}}\right]+25
\end{equation}
The best fit of distance modulus as a function $\mu(z)$ of
redshift $z$ for our theoretical model and the Supernova Type Ia
Union2 sample are drawn in figure 10 for our best fit values of
$A$, $B$ with the other previously chosen parameters. In Figure
11, we have shown that the variation of the curves with slightly
changes in the value of $A$ and $B$ ($A=0.001 \& B=0.13$ for Black
line; $A=0.03 \& B=0.05$ for Red line;$A=0.002 \& B=0.025$ for
Green line). From the curves, we see that the theoretical GCCG
with LQC is in agreement with the union2 sample data.
\begin{figure}
\includegraphics[height=2in]{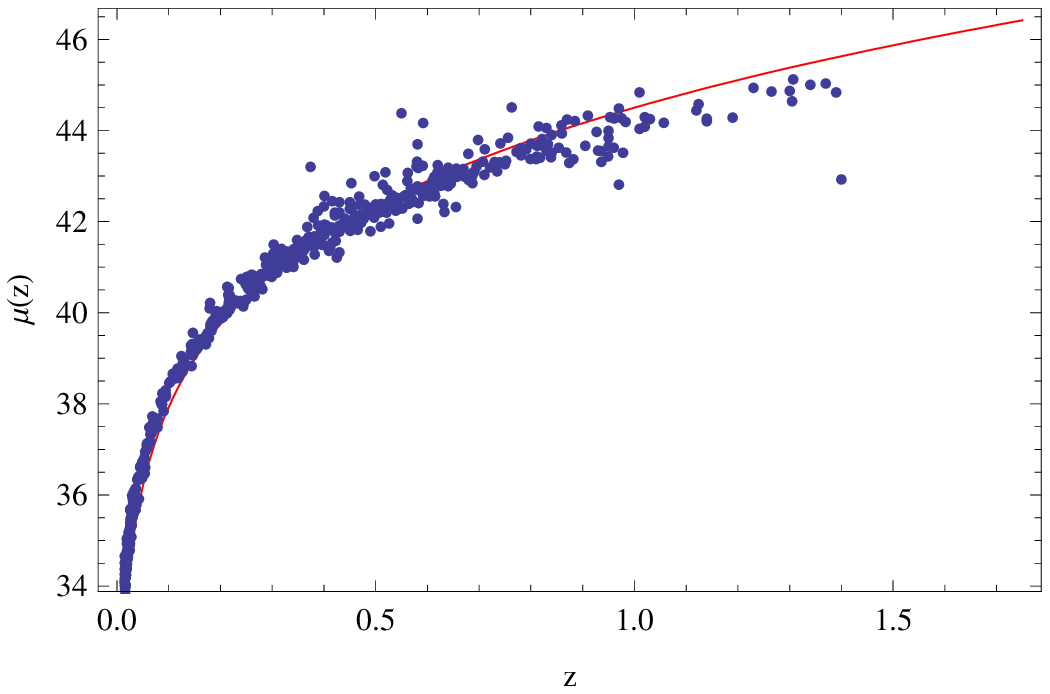}~~~~
\includegraphics[height=2in]{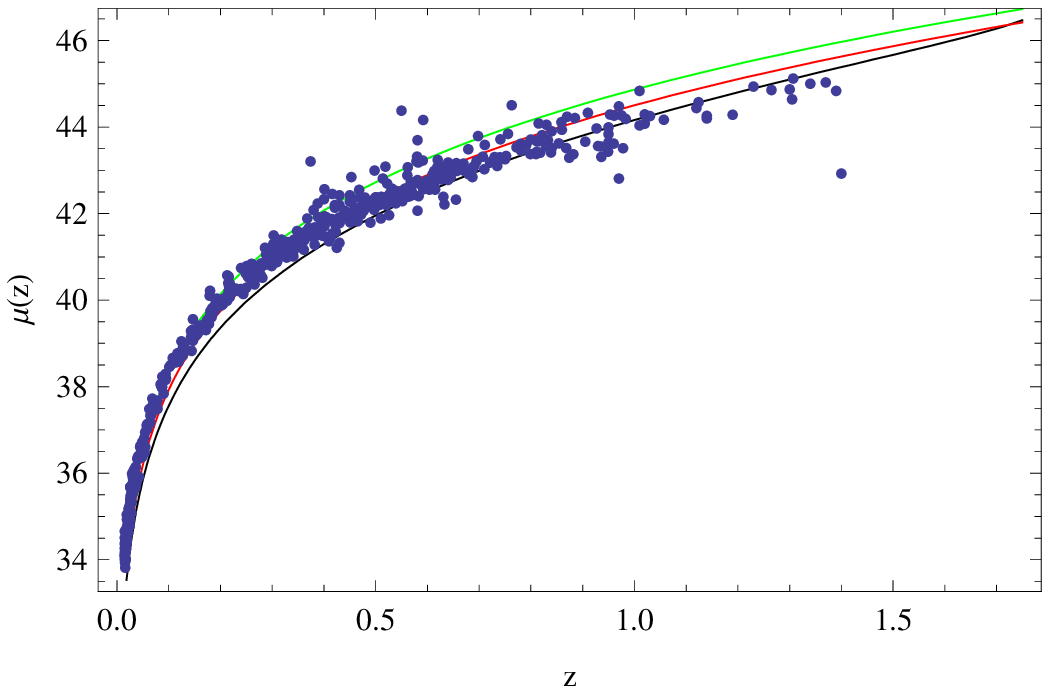}~~\\
\vspace{1mm}
~~~~~~~~~~~~~~~~~~~~~~~~Fig.10~~~~~~~~~~~~~~~~~~~~~~~~~~~~~~~~~~~~~~~~~~~~~~~~~~~~~~Fig.11~~\\
\vspace{2mm} Fig.10 show $\mu(z)$ vs $z$ for our GCCG with LQC
(solid red line) and the Union2 sample (dotted points). Fig 11
shows the same for different value of $A$ \& $B$ ($A=0.001 \&
B=0.13$ for Black line; $A=0.03 \& B=0.05$ for Red line;$A=0.002
\& B=0.025$ for Green line). \vspace{1mm}
\end{figure}

\section{Analysis with Supernovae Type Ia $292$ Data [ From
\cite{Riess1,Riess2,Astier}]} In this section we analyzed our GCCG
with LQC model with same spirit (as stated before in the previous
sections) and obtaining the bounds of the arbitrary parameters
($A~ \&~B$) by fixing the cosmological parameters around their
favorable value with the help of observational 292 Supernovae Type
Ia data which belongs to [\cite{Riess1, Riess2,Astier}] and also
shown in Table 5 at Appendix. As like Sec. 3.1, here also we are
applying $\chi^{2}$ minimization technique, where the $\chi^{2}$
statistics is as follows:
\begin{equation}
{\chi}_{(SNeTypeIa)}^{2}=\sum\frac{(H(z)-H_{obs}(z))^{2}}{\sigma^{2}(z)}
\end{equation}
where the $H_{obs}(z)$ and $\sigma(z)$ are given in Table 5 and
also the probability distribution function can be expressed as
\begin{equation}
L= \int e^{-\frac{1}{2}{\chi}_{(SNeTypeIa)}^{2}}P(H_{0})dH_{0}
\end{equation}
 where $P(H_{0})$ is the prior distribution function for $H_{0}$.
By using $\chi^{2}$ minimization technique, here we plot the graph
of the unknown parameters $A$ and $B$ for same values of $\alpha$
(as stated above) and fixing the other parameters for their most
suitable values and draw for different confidence levels (as 66\%,
90\% and 99\%). The best fit values of the parameters $A$ and $B$
are written in Table 6. It is to be noted that our best fit
analysis with SNe Type Ia observational 292 data also support the
theoretical range of the parameters. It is also to be observed
that for different $\alpha(=0.0020,0.0010 \& 0.0005)$ the best fit
value of $A$ and $B$ are almost same but the value of
$\chi_{min}^{2}$ are different for each cases.\\The 66\% (solid,
blue), 90\% (dashed, red) and 99\% (dashed, black) contours for
$(A,B)$ are plotted in figures 12, 13 and 14 for different values
of $\alpha$. Also the best fit values of $A$ and $B$ are tabulated
in Table 6.
\begin{figure}
\includegraphics[height=2in]{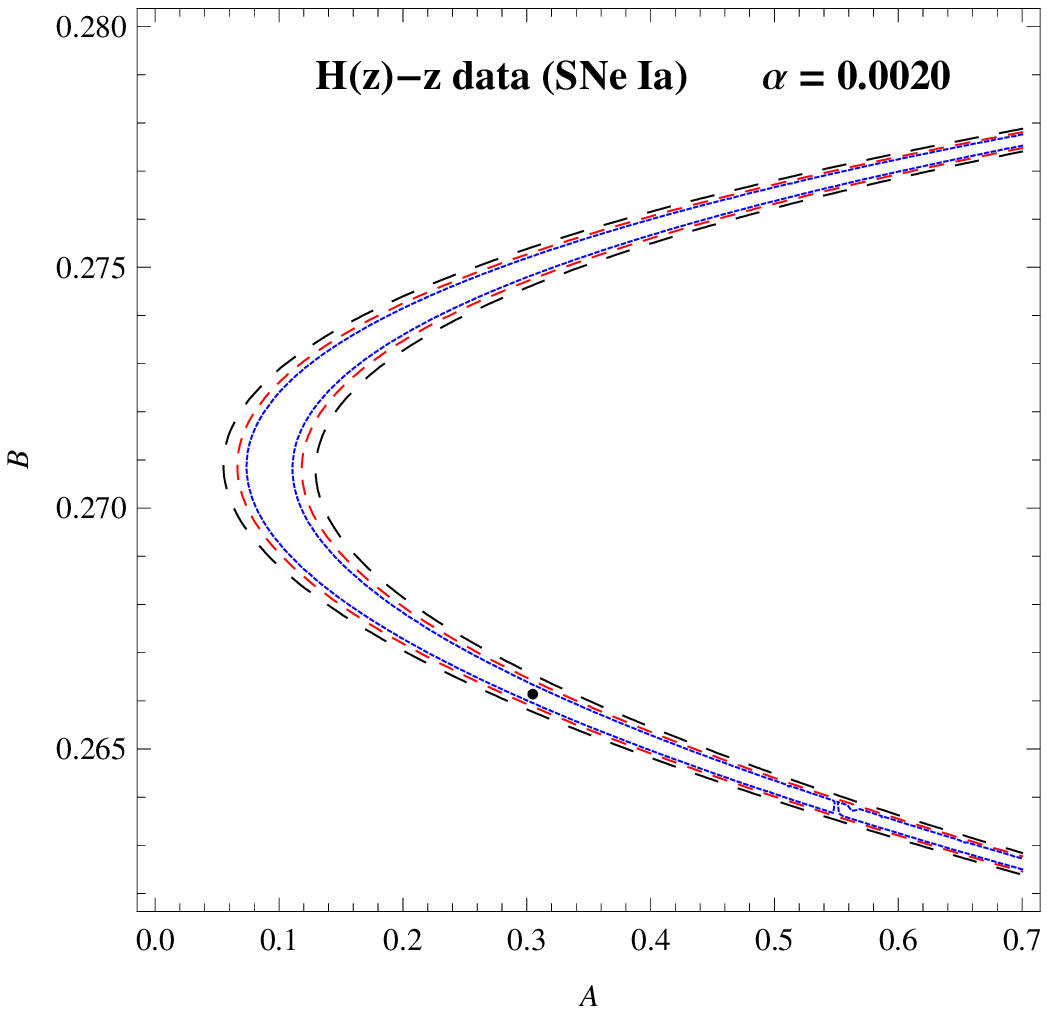}~~
\includegraphics[height=2in]{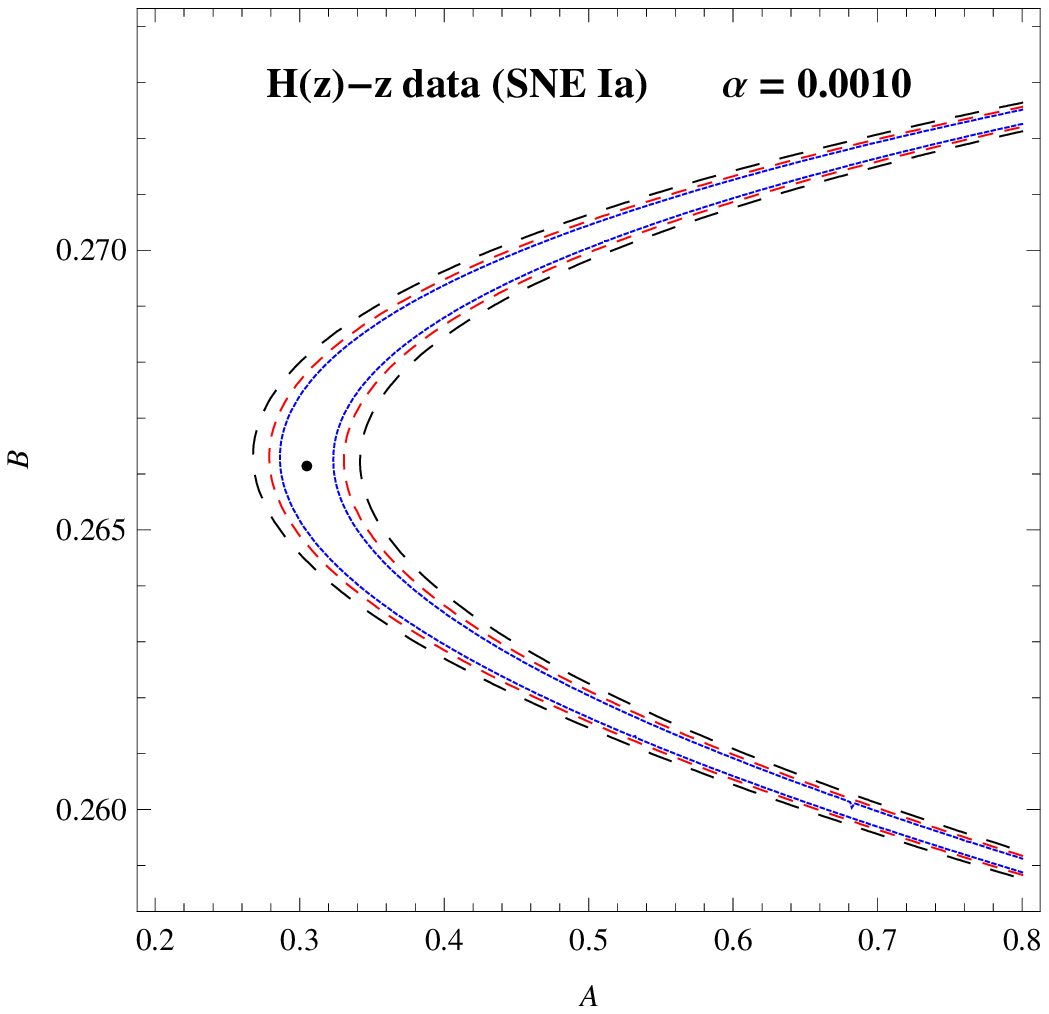}~~
\includegraphics[height=2in]{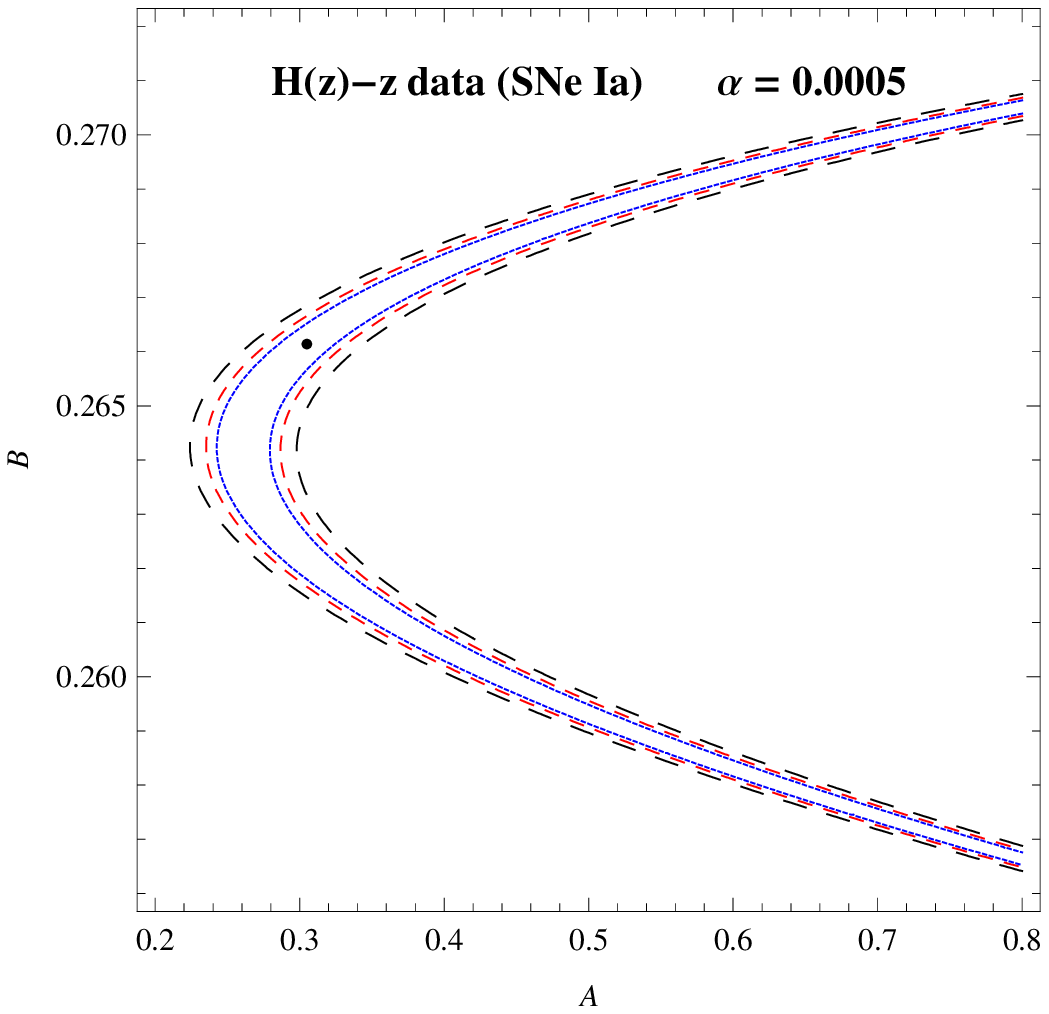}\\
\vspace{1mm}
~~~~~~~~~~~~~~~Fig.12~~~~~~~~~~~~~~~~~~~~~~~~~~~~~~~~~Fig.13~~~~~~~~~~~~~~~~~~~~~~~~~~~~~~~~~~Fig.14~~~~~~\\
\vspace{2mm} Fig.12-14 show that the variation of $A$ with $B$ for
$\Omega_{m0}=6.43\times 10^{-8}, w_{m}=0.051, \omega=-0.92$ with
$\alpha =$ $0.0020$, $0.0010$ \& $0.0005$ respectively for
different confidence levels. The 66\% (solid, blue), 90\% (dashed,
red) and 99\% (dashed, black) contours are plotted in these
figures for the $H(z)$-$z$ of SNe Type Ia data. \vspace{1mm}
\end{figure}
\[
\begin{tabular}{|c|c|c|c|}
\hline
 ~~~~~~$\alpha$ ~~~~&~~~~$A$ ~~~~~& ~~~~~~~$B$ ~~~~~~~~& ~~~$\chi^{2}_{min}$~~~~~~\\
  \hline
  0.0020 & 0.304936 & 0.266141 & 87260.15 \\
  0.0010 & 0.304936 & 0.266141 & 86928.93 \\
  0.0005 & 0.304936 & 0.266141 & 86863.26 \\
  \hline
\end{tabular}
\]
{\bf Table 6:} $H(z)$-$z$ SNe Type Ia : The best fit values of
$A$, $B$ and the minimum values of $\chi^{2}$ for different values
of $\alpha$ and fixed value of other parameters.

\subsection{Redshift-Magnitude Observational Analysis with Supernovae
Type Ia $292$ Data [ From \cite{Riess1,Riess2,Astier}]} In this
subsection we measure the distance modulus (as like Sec. 3.4) of
SNe Type Ia 292 data which belongs to [\cite{Riess1,Riess2,
Astier}]. Here also we use the same luminosity distance $d_{L}(z)$
which is defined as
\begin{equation}
d_{L}(z)=(1+z)H_{0}\int_{0}^{z}\frac{dz'}{H(z')}
\end{equation}
and the distance modulus for SNe Type Ia observational 292 data is
given below:
\begin{equation}
\mu(z)=5\log_{10} \left[\frac{d_{L}(z)/H_{0}}{1~
\text{MPc}}\right]+25
\end{equation}
As stated above, the best fit of distance modulus $\mu(z)$ which
is a function of redshift $z$ for our theoretical model and the
SNe Type Ia 292 data from [\cite{Riess1,Riess2,Astier}] are drawn
in figure 15 with most favorable different values of $A$, $B$ with
the previously chosen other parameters. From the curves, we can
conclude that the theoretical GCCG with LQC is in agreement with
the SNe Type Ia 292 data from [\cite{Riess1,Riess2,Astier}].
\begin{figure}
~~~~~~~~~~~~~~~~~~~~~~~~~
\includegraphics[height=2.8in]{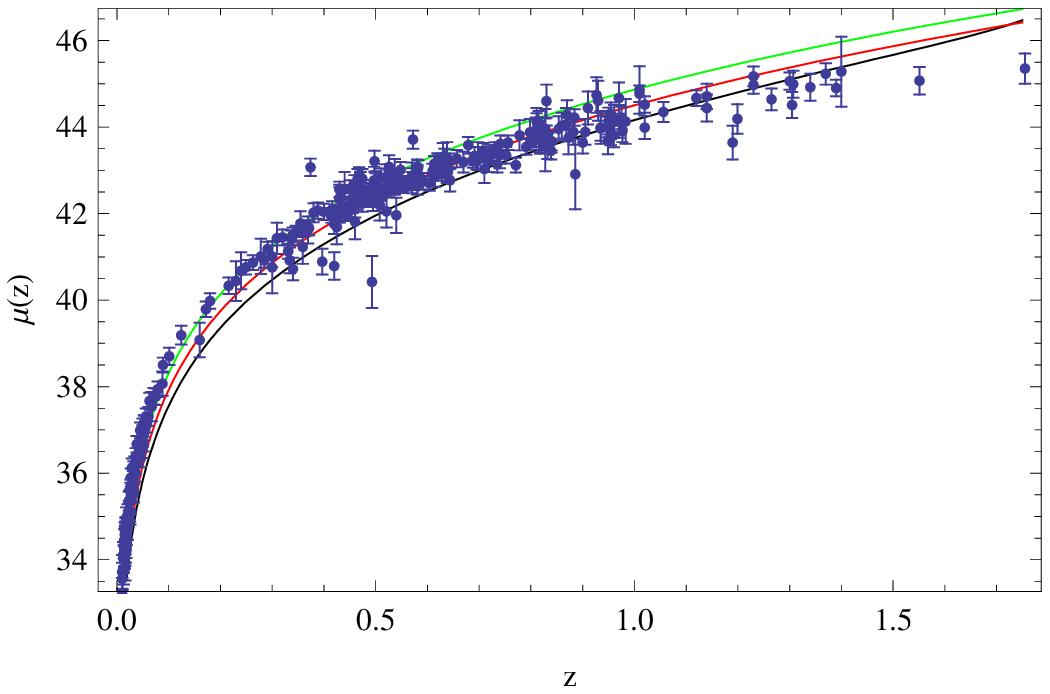}\\
\vspace{1mm}
~~~~~~~~~~~~~~~~~~~~~~~~~~~~~~~~~~~~~~~~~~~~~~~~~~~~~~~~~~~~~~~~~~Fig.15~~~~~~~~\\
\vspace{1mm} Fig.15 shows $\mu(z)$ vs $z$ for our GCCG with LQC
for SNe Type Ia data(dotted points) Three lines are drawn for
different value of A \& B ($A=0.001 \& B=0.13$ for Black line;
$A=0.03 \& B=0.05$ for Red line;$A=0.002 \& B=0.025$ for Green
line). \vspace{1mm}
\end{figure}

\section{\normalsize\bf{Study of Future Singularities}}
In recent time, the well established universal fact for any energy
dominated model of the universe is intended to the result in
future singularity. Without studying of these singularities, the
ultimate goal of this study of our model become incomplete. A well
known cosmological hypothesis is that the universe dominated by
phantom energy ends with a future singularity, which violates the
dominant energy condition (DEC), known as Big
Rip[\cite{Caldwell2003}]. In 2005, \cite{Nojiri2005} studied the
various types of singularities for an phantom energy dominated
universe. There are many effective approaches have been adopted by
some authors [\cite{Sami, Naskar, Samart,Cailleteau et al., Singh,
Corichi, Lamon, Singh et al.}] to study the future singularities.
Singularities are basically characterized by the growth of energy
and curvature at the time of occurrence of them. It is observed
that the quantum effects are not only very dominant near the
singularities , they may prevent these singularities. All
different types of future singularities for different scenario was
discussed by \cite{Nojiri2011}. Recently \cite{Bamba} studied
future singularities in the context of LQC and shown that some of
these singularities may be avoided. In this regards some works
have been done by \cite{Rudra2012a,Ratul,Rudra2012b}. Future
singularities are basically classified in four types and in each
cases our model have been tested for those scenarios as follows:

$\bullet~~${\normalsize\bf{TYPE-I Singularity (Big Rip):}}When
$\rho\rightarrow\infty$ and $|p|\rightarrow\infty$ for
 $a\rightarrow \infty$ and $t\rightarrow t_s$.

In this present scenario our predicted model of LQC with GCCG and
DM in non interacting scenario have been tested and we have
\begin{eqnarray*}
a\rightarrow\infty:
\begin{array}{ll}
\rho_{x}\rightarrow\infty ~~~~~~~~~~~~\text{for}~~ 1+\omega<0
\Rightarrow |p_{x}|\rightarrow 0\\
\rho_{x}\rightarrow (C+1)^{\frac{1}{\alpha+1}}~~\text{for}~~
1+\omega>0 \Rightarrow |p_{x}|\rightarrow
(C+1)^{\frac{1}{\alpha+1}}
\end{array}
\end{eqnarray*}
and from the above results we can conclude that there is no
possibility of Type-I i.e., ``Big Rip'' singularity and the result
is absolutely accordance with the work of some authors
[\cite{Gon,Bamba,Ratul}] who have shown that ``Big Rip'' can be
easily avoided in LQC with non interacting GCCG and DM and
produced a singularity free late universe.

$\bullet~~${\normalsize\bf{TYPE-II Singularity (Sudden):}} When
$\rho\rightarrow\rho_s$ and $|p|\rightarrow\infty$ for
 $a\rightarrow a_s$ and $t\rightarrow t_s$.

In this case we have been again considering our predicted model of
LQC with non interacting GCCG and DM and we find that
\begin{eqnarray*}
a\rightarrow a_{s}\sim 0:
\begin{array}{ll}
\rho_{x}\rightarrow\infty ~~~~~~~~~~~~~\text{for}~~ 1+\omega>0
\Rightarrow |p_{x}|\rightarrow 0\\
\rho_{x}\rightarrow (C+1)^{\frac{1}{\alpha+1}}~~\text{for}~~
1+\omega<0 \Rightarrow |p_{x}|\rightarrow
(C+1)^{\frac{1}{\alpha+1}}
\end{array}
\end{eqnarray*}
and it can be concluded that there is no possibility of the
Type-II or ``Sudden'' singularity for our predicted model.

$\bullet~~${\normalsize\bf{TYPE-III Singularity (Big Freeze):}}
When $\rho\rightarrow\infty$ and $|p|\rightarrow\infty$ if
 $a\rightarrow a_s$ and $t\rightarrow t_s$.

In this present condition, it can be quite evidently concluded
from our model of LQC with non interacting GCCG and DM that it
does not support this Type-III or ``Big Freeze'' singularity. In
this regards there are some works by some authors [\cite{Ratul,
Rudra2012b}] in supports of this result.

$\bullet~~${\normalsize\bf{TYPE-IV Singularity (Generalized
Sudden):}} For $t\rightarrow t_s$, $a\rightarrow a_s$,
$\rho\rightarrow 0$ and $|p|\rightarrow 0$

In this regards we have expressed scale factor $a(t)$ in terms of
energy density of GCCG as follows:
\begin{eqnarray*}
a=\left[\frac{B}{(\rho_{x}^{ \alpha+1}-C)^{\omega+1}-1}\right]
^{\frac{1}{3(1+\alpha)(1+\omega)}}
\end{eqnarray*}
and therefore it can be easily concluded that this type of
singularity is not supported by our predicted LQC model with non
interacting GCCG and DM.

\section{\normalsize\bf{Discussions}}
We have assumed the FRW universe in loop quantum cosmology (LQC)
model filled with the dark matter and the Generalized Cosmic
Chaplygin gas (GCCG) type dark energy. We present the Hubble
parameter $H(z)$ in terms of the observable parameters
$\Omega_{m0}$ and $H_{0}$ with the redshift $z$ and the other
parameters like $A$, $B$, $w_{m}$, $ \omega$ and $\alpha$. From
Stern data set (12 points), we have obtained the best fit values
of two arbitrary parameters $(A,B)$ in table 2 by fixing other
parameters $\Omega_{m0}=0.0643,w_{m}=0.051,\omega=-0.92$ and
$\alpha=0.002,0.001,0.0005$ by minimizing the $\chi^{2}$ test. The
bounds of the parameters $(A,B)$ are obtained by 66\%, 90\% and
99\% confidence levels in figures 1-3. Next due to joint analysis
with BAO and CMB observations, we have also obtained the best fit
values of the parameters ($A,B$) by fixing the other parameters
(same values) in tables 3 and 4 respectively. Also the bounds of
the parameters $(A,B)$ due to joint analysis with BAO and CMB
observations are obtained by 66\%, 90\% and 99\% confidence levels
in figures 4-6 and figures 7-9 respectively. From tables 1-3, we
see that when $\alpha=0.002,0.001$, the best-fit values of $A$ and
$B$ are positive for all our observational data. But if
$\alpha=0.0005$, the best-fit value of $A$ is negative and
best-fit value of $B$ is still positive for all our observational
data. From the best fit value of distance modulus $\mu(z)$ for our
theoretical GCCG model in LQC is drawn in figure 10. Fig 11 shows
the same for different value of $A$ \& $B$ ($A=0.001 \& B=0.13$
for Black line; $A=0.03 \& B=0.05$ for Red line;$A=0.002 \&
B=0.025$ for Green line). From the figure, we have concluded that
our predicted theoretical GCCG model in LQC permitted the union2
sample data sets of SNe Type Ia. After that we have considered SNe
Type Ia Riess 292 data from \cite{Riess1,Riess2, Astier} and
tested our theoretical GCCG with LQC model by minimizing the
$\chi^{2}$ test (same as stated above) and obtained the bounds of
the parameters $(A,B)$ given in Table 6. When the other parameters
are fixed at their suitable value as $\Omega_{m0}=6.43\times
10^{-8}, w_{m}=0.051, \omega=-0.92$ with $\alpha =$ $0.0020$,
$0.0010$ \& $0.0005$, we have drawn the figure 12-14 respectively
and the best fitted values of $A=0.304936$ \& $B=0.266141$ are
almost same in every cases of $\alpha$. Figure 15 shows the
distance modulus $\mu(z)$ of our theoretical GCCG model in LQC
together with Riess 292 data for different favorable values of $A$
\& $B$ ($A=0.001 \& B=0.13$ for Black line; $A=0.03 \& B=0.05$ for
Red line;$A=0.002 \& B=0.025$ for Green line) vs redshift $z$ and
which depicted that our theoretical GCCG model in LQC suitably
permitted with the Riess 292 data of SNe Type Ia. From the above
results, we can finally conclude that our theoretical GCCG model
in LQC is in agreement with the Supernovae Type Ia sample data. In
addition, we have also investigated in details about the Future
Singularities like Type-I, Type-II, Type-III and Type-IV that may
be formed and or avoided in this model and it is found that our
model is completely free from any types of future singularities.

\section*{Acknowledgements}
The authors are thankful to IUCAA, Pune, India for warm
hospitality where part of the work was carried out. The author
(UD) is thankful to CSIR, Govt. of India for providing research
project grant (No. 03(1206)/12/EMR-II).

\appendix
\section{Appendix material}
\clearpage
\begin{deluxetable}{@{}llcrrrrrrrrrrr@{}}
\tablecolumns{5}\tablewidth{0pt} \tablenum{5}
\tabletypesize{\scriptsize}\tablecaption{SNe la 292 Data Set from
Riess et al. (2004, 2007); Astier et al. (2006)} \tablehead{
\colhead{Name} & \colhead{$z$} & \colhead{$H(z)$} &
\colhead{$\sigma(z)$} & \colhead{Type}} \startdata
  SN90O  & 0.030  &  35.90  & 0.21  &   Gold \\
  SN90T  & 0.040  &  36.38  & 0.20  &   Gold \\
  SN90af &  0.050 &  36.84  & 0.22  &   Gold \\
  SN91ag & 0.014  &  34.13  & 0.29  &   Gold \\
  SN91U  & 0.033  &  35.53  & 0.21  &   Gold \\
  SN91S  & 0.056  &  37.31  & 0.19  &   Gold \\
  SN92al & 0.014  &  34.12  & 0.29  &   Gold \\
  SN92bo & 0.017  &  34.70  & 0.26  &   Gold \\
  SN92bc & 0.018  &  34.96  & 0.25  &   Gold \\
  SN92ag & 0.026  &  35.06  & 0.25  &   Silver \\
  SN92P  & 0.026  &  35.63  & 0.22  &   Gold \\
  SN92bg & 0.036  &  36.17  & 0.20  &   Gold \\
  SN92bl &  0.043 &  36.52  & 0.19  &   Gold \\
  SN92bh &  0.045 &  36.99  & 0.18  &   Gold \\
  SN92J  & 0.046  &  36.35  & 0.21  &   Gold \\
  SN92bk & 0.058  &  37.13  & 0.19  &   Gold \\
  SN92au & 0.061  &  37.31  & 0.22  &   Gold \\
  SN92bs & 0.063  &  37.67  & 0.19  &   Gold \\
  SN92ae & 0.075  &  37.77  & 0.19  &   Gold \\
  SN92bp & 0.079  &  37.94  & 0.18  &   Gold \\
  SN92br & 0.088  &  38.07  & 0.28  &   Gold \\
  SN92aq & 0.101  &  38.70  & 0.20  &   Gold \\
  SN93ae & 0.018  &  34.29  & 0.25  &   Gold \\
  SN93H  & 0.025  &  35.09  & 0.22  &   Gold \\
  SN93ah & 0.028  &  35.53  & 0.22  &   Gold \\
  SN93ac & 0.049  &  36.90  & 0.21  &   Silver \\
  SN93ag & 0.050  &  37.07  & 0.19  &   Gold \\
  SN93O  & 0.052  &  37.16  & 0.18  &   Gold \\
  SN93B  & 0.071  &  37.78  & 0.19  &   Gold \\
  SN94S  & 0.016  &  34.50  & 0.27  &   Gold \\
  SN94M  & 0.024  &  35.09  & 0.22  &   Gold \\
  SN94Q  & 0.029  &  35.70  & 0.21  &   Gold \\
  SN94T  & 0.036  &  36.01  & 0.21  &   Gold \\
  SN94C  & 0.051  &  36.67  & 0.17  &   Silver \\
  SN94B  & 0.089  &  38.50  & 0.17  &   Silver \\
  SN95K  & 0.478  &  42.48  & 0.23  &   Gold \\
  SN95ak & 0.021  &  34.70  & 0.24  &   Silver \\
  SN95E  & 0.011  &  32.95  & 0.35  &   Silver \\
  SN95bd & 0.015  &  34.07  & 0.28  &   Silver \\
  SN95ac & 0.049  &  36.55  & 0.20  &   Gold \\
  SN95ar & 0.465  &  42.81  & 0.22  &   Silver \\
  SN95as & 0.498  &  43.21  & 0.24  &   Silver \\
  SN95aw & 0.400  &  42.04  & 0.19  &   Gold \\
  SN95ax & 0.615  &  42.85  & 0.23  &   Gold \\
  SN95ay & 0.480  &  42.37  & 0.20  &   Gold \\
  SN95az & 0.450  &  42.13  & 0.21  &   Gold \\
  SN95ba & 0.388  &  42.07  & 0.19  &   Gold \\
  SN95M  & 0.053  &  37.17  & 0.16  &   Silver \\
  SN95ae & 0.067  &  37.54  & 0.34  &   Silver \\
  SN95ao & 0.300  &  40.76  & 0.60  &   Silver \\
  SN95ap & 0.230  &  40.44  & 0.46  &   Silver \\
  SN96E  & 0.425  &  41.69  & 0.40  &   Gold \\
  SN96H  & 0.620  &  43.11  & 0.28  &   Gold \\
  SN96I  & 0.570  &  42.80  & 0.25  &   Gold \\
  SN96J  & 0.300  &  41.01  & 0.25  &   Gold \\
  SN96K  & 0.380  &  42.02  & 0.22  &   Gold \\
  SN96U  & 0.430  &  42.33  & 0.34  &   Gold \\
  SN96Z  & 0.008  &  32.45  & 0.45  &   Silver \\
  SN96bo & 0.016  &  33.82  & 0.30  &   Silver \\
  SN96bv & 0.016  &  34.21  & 0.27  &   Silver \\
  SN96bk & 0.007  &  32.09  & 0.53  &   Silver \\
  SN96C  & 0.027  &  35.90  & 0.21  &   Gold \\
  SN96bl & 0.034  &  36.19  & 0.20  &   Gold \\
  SN96ab & 0.124  &  39.19  & 0.22  &   Gold \\
  SN96cf & 0.570  &  42.77  & 0.19  &   Silver \\
  SN96cg & 0.490  &  42.58  & 0.19  &   Silver \\
  SN96ci & 0.495  &  42.25  & 0.19  &   Gold \\
  SN96cl & 0.828  &  43.96  & 0.46  &   Gold \\
  SN96cm & 0.450  &  42.58  & 0.19  &   Silver \\
  SN96cn & 0.430  &  42.56  & 0.18  &   Silver \\
  SN96V  & 0.024  &  35.33  & 0.26  &   Silver \\
  SN96T  & 0.240  &  40.68  & 0.43  &   Silver \\
  SN96R  & 0.160  &  39.08  & 0.40  &   Silver \\
  SN97eq & 0.538  &  42.66  & 0.18  &   Gold \\
  SN97ek & 0.860  &  44.03  & 0.30  &   Gold \\
  SN97ez & 0.778  &  43.81  & 0.35  &   Gold \\
  SN97as & 0.508  &  42.19  & 0.35  &   Gold \\
  SN97aw & 0.440  &  42.56  & 0.40  &   Silver \\
  SN97bb & 0.518  &  42.83  & 0.31  &   Gold \\
  SN97bh & 0.420  &  41.76  & 0.23  &   Silver \\
  SN97bj & 0.334  &  40.92  & 0.30  &   Gold \\
  SN97ce & 0.440  &  42.07  & 0.19  &   Gold \\
  SN97cj & 0.500  &  42.73  & 0.20  &   Gold \\
  SN97do & 0.010  &  33.72  & 0.39  &   Gold \\
  SN97E  & 0.013  &  34.02  & 0.31  &   Gold \\
  SN97Y  & 0.016  &  34.53  & 0.27  &   Gold \\
  SN97cn & 0.017  &  34.71  & 0.28  &   Gold \\
  SN97dg & 0.029  &  36.13  & 0.21  &   Gold \\
  SN97F  & 0.580  &  43.04  & 0.21  &   Gold \\
  SN97H  & 0.526  &  42.56  & 0.18  &   Gold \\
  SN97I  & 0.172  &  39.79  & 0.18  &   Gold \\
  SN97N  & 0.180  &  39.98  & 0.18  &   Gold \\
  SN97O  & 0.374  &  43.07  & 0.20  &   Silver \\
  SN97P  & 0.472  &  42.46  & 0.19  &   Gold \\
  SN97Q  & 0.430  &  41.99  & 0.18  &   Gold \\
  SN97R  & 0.657  &  43.27  & 0.20  &   Gold \\
  SN97ac & 0.320  &  41.45  & 0.18  &   Gold \\
  SN97af & 0.579  &  42.86  & 0.19  &   Gold \\
  SN97ai & 0.450  &  42.10  & 0.23  &   Gold \\
  SN97aj & 0.581  &  42.63  & 0.19  &   Gold \\
  SN97am & 0.416  &  42.10  & 0.19  &   Gold \\
  SN97ap & 0.830  &  43.85  & 0.19  &   Gold \\
  SN97ck & 0.970  &  44.13  & 0.38  &   Silver \\
  SN98ax & 0.497  &  42.77  & 0.31  &   Silver \\
  SN98aw & 0.440  &  42.02  & 0.19  &   Silver \\
  SN98ay & 0.638  &  43.29  & 0.36  &   Silver \\
  SN98ba & 0.430  &  42.36  & 0.25  &   Gold \\
  SN98be & 0.644  &  42.77  & 0.26  &   Silver \\
  SN98as & 0.355  &  41.77  & 0.28  &   Silver \\
  SN98bi & 0.740  &  43.35  & 0.30  &   Gold \\
  SN98ac & 0.460  &  41.81  & 0.40  &   Gold \\
  SN98M  & 0.630  &  43.26  & 0.37  &   Gold \\
  SN98J  & 0.828  &  43.59  & 0.61  &   Gold \\
  SN98I  & 0.886  &  42.91  & 0.81  &   Silver \\
  SN98bp & 0.010  &  33.20  & 0.38  &   Gold \\
  SN98ef & 0.017  &  34.18  & 0.26  &   Gold \\
  SN98V  & 0.017  &  34.47  & 0.26  &   Gold \\
  SN98co & 0.017  &  34.62  & 0.27  &   Gold \\
  SN98eg & 0.023  &  35.35  & 0.22  &   Gold \\
  SN98cs & 0.032  &  36.08  & 0.20  &   Gold \\
  SN98dx & 0.053  &  36.95  & 0.19  &   Gold \\
  SN99Q2 & 0.459  &  42.67  & 0.22  &   Gold \\
  SN99U2 & 0.511  &  42.83  & 0.21  &   Gold \\
  SN99S  & 0.474  &  42.81  & 0.22  &   Gold \\
  SN99N  & 0.537  &  42.85  & 0.41  &   Gold \\
  SN99M  & 0.493  &  40.42  & 0.60  &   Silver \\
  SN99fn & 0.477  &  42.38  & 0.21  &   Gold \\
  SN99ff & 0.455  &  42.29  & 0.28  &   Gold \\
  SN99fj & 0.815  &  43.75  & 0.33  &   Gold \\
  SN99fm & 0.949  &  44.00  & 0.24  &   Gold \\
  SN99fk & 1.056  &  44.35  & 0.23  &   Gold \\
  SN99fw & 0.278  &  41.01  & 0.41  &   Gold \\
  SN99cp & 0.010  &  33.56  & 0.37  &   Gold \\
  SN99dq & 0.013  &  33.73  & 0.30  &   Gold \\
  SN99dk & 0.014  &  34.43  & 0.29  &   Gold \\
  SN99aa & 0.015  &  34.58  & 0.28  &   Gold \\
  SN99X  & 0.025  &  35.40  & 0.22  &   Gold \\
  SN99gp & 0.026  &  35.57  & 0.21  &   Gold \\
  SN99cc & 0.031  &  35.84  & 0.21  &   Gold \\
  SN99ef & 0.038  &  36.67  & 0.19  &   Gold \\
  SN99fv & 1.199  &  44.19  & 0.34  &   Gold \\
  SN99fh & 0.369  &  41.62  & 0.31  &   Silver \\
  SN99da & 0.012  &  34.05  & 0.36  &   Silver \\
  SN00ec & 0.470  &  42.76  & 0.21  &   Gold \\
  SN00dz & 0.500  &  42.74  & 0.24  &   Gold \\
  SN00ea & 0.420  &  40.79  & 0.32  &   Silver \\
  SN00eg & 0.540  &  41.96  & 0.41  &   Gold \\
  SN00ee & 0.470  &  42.73  & 0.23  &   Gold \\
  SN00eh & 0.490  &  42.40  & 0.25  &   Gold \\
  SN00fr & 0.543  &  42.67  & 0.19  &   Gold \\
  SN00dk & 0.016  &  34.41  & 0.27  &   Gold \\
  SN00B  & 0.019  &  34.59  & 0.25  &   Gold \\
  SN00fa & 0.021  &  35.05  & 0.23  &   Gold \\
  SN00cn & 0.023  &  35.14  & 0.22  &   Gold \\
  SN00bk & 0.026  &  35.35  & 0.23  &   Gold \\
  SN00cf & 0.036  &  36.39  & 0.19  &   Gold \\
  SN00ce & 0.016  &  34.47  & 0.26  &   Silver \\
  SN01iv & 0.397  &  40.89  & 0.30  &   Silver \\
  SN01iw & 0.340  &  40.72  & 0.26  &   Silver \\
  SN01jh & 0.884  &  44.22  & 0.19  &   Gold \\
  SN01hu & 0.882  &  43.89  & 0.30  &   Gold \\
  SN01ix & 0.710  &  43.03  & 0.32  &   Silver \\
  SN01iy & 0.570  &  42.87  & 0.31  &   Gold \\
  SN01jp & 0.528  &  42.76  & 0.25  &   Gold \\
  SN01V  & 0.016  &  34.13  & 0.27  &   Gold \\
  SN01fo & 0.771  &  43.12  & 0.17  &   Gold \\
  SN01fs & 0.873  &  43.75  & 0.38  &   Silver \\
  SN01hs & 0.832  &  43.55  & 0.29  &   Gold \\
  SN01hx & 0.798  &  43.88  & 0.31  &   Gold \\
  SN01hy & 0.811  &  43.97  & 0.35  &   Gold \\
  SN01jb & 0.698  &  43.33  & 0.32  &   Silver \\
  SN01jf & 0.815  &  44.09  & 0.28  &   Gold \\
  SN01jm & 0.977  &  43.91  & 0.26  &   Gold \\
  SN01kd & 0.935  &  43.99  & 0.38  &   Silver \\
  SN02P  & 0.719  &  43.22  & 0.26  &   Silver \\
  SN02ab & 0.422  &  42.02  & 0.17  &   Silver \\
  SN02ad & 0.514  &  42.39  & 0.27  &   Silver \\
  1997ff & 1.755  &  45.35  & 0.35  &   Gold \\
  2002dc & 0.475  &  42.24  & 0.20  &   Gold \\
  2002dd & 0.950  &  43.98  & 0.34  &   Gold \\
  2003aj &  1.307 &  44.99  & 0.31  &    Silver \\
  2002fx &  1.400 &  45.28  & 0.81  &    Silver \\
  2003eq &  0.840 &  43.67  & 0.21  &     Gold \\
  2003es &  0.954 &  44.30  & 0.27  &    Gold \\
  2003az &  1.265 &  44.64 &  0.25   &      Silver \\
  2002kc &  0.216 &  40.33 &  0.19   &   Silver \\
  2003eb &  0.900 &  43.64 &  0.25   &  Gold \\
  2003XX &  0.935 &  43.97 &  0.29   &   Gold \\
  2002hr &  0.526 &  43.08 &  0.27   &   Silver \\
  2003bd &  0.670 &  43.19 &  0.24   & Gold \\
  2002kd &  0.735 &  43.14 &  0.19   &   Gold \\
  2003be &  0.640 &  43.01 &  0.25   &   Gold \\
  2003dy &  1.340 &  44.92 &  0.31   &   Gold \\
  2002ki &  1.140 &  44.71 &  0.29  &   Gold \\
  2003ak &  1.551 &  45.07 &  0.32   &  Silver \\
  2002hp &  1.305 &  44.51 &  0.30   &  Gold \\
  2002fw &  1.300 &  45.06 &  0.20   &   Gold \\
 HST04Pat &  0.970 &  44.67 &  0.36  &    Gold \\
 HST04Mcg &  1.370 &  45.23 &  0.25  &   Gold \\
 HST05Fer &  1.020 &  43.99 &  0.27  &    Gold \\
 HST05Koe &  1.230 &  45.17 &  0.23  &   Gold \\
 HST05Dic &  0.638 &  42.89 &  0.18  &    Silver \\
 HST04Gre &  1.140 &  44.44 &  0.31  &   Gold \\
 HST04Omb &  0.975 &  44.21 &  0.26  &    Gold \\
 HST05Red &  1.190 &  43.64 &  0.39  &  Silver \\
 HST05Lan &  1.230 &  44.97 &  0.20  &    Gold \\
 HST04Tha &  0.954 &  43.85 &  0.27  &    Gold \\
 HST04Rak &  0.740 &  43.38 &  0.22  &    Gold \\
 HST05Zwi &  0.521 &  42.05  & 0.37  &    Silver \\
 HST04Hawk &  0.490 &  42.54 &  0.24 &     Silver \\
 HST04Kur &  0.359 &  41.23 &  0.39  &    Silver \\
 HST04Yow  & 0.460 &  42.23 &  0.32  &    Gold \\
 HST04Man &  0.854 &  43.96 &  0.29  &    Gold \\
 HST05Spo &  0.839 &  43.45 &  0.20  &    Gold \\
 HST04Eag  & 1.020 &  44.52 &  0.19  &    Gold \\
 HST05Gab &  1.120 &  44.67 &  0.18  &    Gold \\
 HST05Str &  1.010 &  44.77 &  0.19  &   Gold \\
 HST04Sas  & 1.390 &  44.90 &  0.19  &   Gold \\
 SN88U  & 0.309 &  41.43 &  0.36  &    Silver \\
 SN-03D1au &  0.504 & 42.61 & 0.17  &    Gold \\
 SN-03D1aw &  0.582 &  43.07 &  0.17 & Gold \\
 SN-03D1ax & 0.496 &  42.36 &  0.17  &    Gold \\
 SN-03D1bp & 0.346 & 41.55 & 0.17 & Silver \\
 SN-03D1cm &  0.870 &  44.28 & 0.34 &      Gold \\
 SN-03D1co & 0.679 &  43.58 &  0.19 &     Gold \\
 SN-03D1ew &  0.868 & 44.06 &0.38& Silver \\
 SN-03D1fc &  0.331 & 41.13 &  0.17 &     Gold \\
 SN-03D1fl & 0.688 &  43.23 &  0.17& Gold \\
 SN-03D1fq &  0.800 &   43.67 & 0.19& Gold \\
 SN-03D1gt & 0.548 &   43.01 &   0.18 &     Silver \\
 SN-03D3af & 0.532 & 42.78 & 0.18&      Gold \\
 SN-03D3aw &  0.449 &   42.05 & 0.17& Gold \\
 SN-03D3ay & 0.371 &  41.67 &   0.17&      Gold \\
 SN-03D3ba & 0.291 & 41.18 & 0.17& Silver \\
 SN-03D3bh &  0.249 &   40.76 &   0.17 &  Gold \\
 SN-03D3cc & 0.463 & 42.27 &   0.17 &       Gold \\
 SN-03D3cd & 0.461 & 42.22 &  0.17 &  Gold \\
 SN-03D4ag &  0.285 &   40.92 &    0.17 &  Gold \\
 SN-03D4at & 0.633 & 43.32 &  0.18 &  Gold \\
 SN-03D4aud & 0.468 & 42.89 &    0.18 &  Silver \\
 SN-03D4bcd & 0.572 & 43.71 &  0.21 &  Silver \\
 SN-03D4cn & 0.818 & 43.72 &  0.34 &  Silver \\
 SN-03D4cx & 0.949 & 43.69 &    0.32 &  Gold \\
 SN-03D4cy & 0.927 & 44.74 &  0.41 &  Silver \\
 SN-03D4cz  & 0.695 & 43.21 &  0.19 &  Gold \\
 SN-03D4dh & 0.627 & 42.93 &  0.17 &  Gold \\
 SN-03D4di & 0.905 &   43.89 &  0.30 &  Gold \\
 SN-03D4dy & 0.604 & 42.70 &    0.17 &  Gold \\
 SN-03D4fd & 0.791 & 43.54 &  0.18 &  Gold \\
 SN-03D4gf & 0.581 & 42.95  &   0.17 &  Silver \\
 SN-03D4gg & 0.592 & 42.75  & 0.19 &  Gold \\
 SN-03D4gl & 0.571 &   42.65 &  0.18 &  Gold \\
 SN-04D1ag & 0.557 & 42.70 &  0.17 &  Gold \\
 SN-04D1aj & 0.721 & 43.39  & 0.20 &  Silver \\
 SN-04D1ak & 0.526 & 42.83 &  0.17 &  Silver \\
 SN-04D2cf & 0.369 & 41.67 &  0.17  & Gold \\
 SN-04D2fp & 0.415 &   41.96 &  0.17 &  Gold \\
 SN-04D2fs & 0.357 & 41.63 &  0.17  & Gold \\
 SN-04D2gb & 0.430 &   41.96 &  0.17 &  Gold \\
 SN-04D2gc & 0.521 & 42.62 &  0.17 &  Silver \\
 SN-04D2gp & 0.707 & 43.42  & 0.21 &  Gold \\
 SN-04D2iu & 0.691 &   43.33 &    0.21  & Silver \\
 SN-04D2ja & 0.741 & 43.61 &  0.20 &  Silver \\
 SN-04D3co & 0.620 & 43.21  & 0.18 &  Gold \\
 SN-04D3cp & 0.830 & 44.60 &  0.38 &  Silver \\
 SN-04D3cy & 0.643 & 43.21 &    0.18 &  Gold \\
 SN-04D3dd & 1.010 & 44.86 &  0.55 &  Silver \\
 SN-04D3df & 0.470 & 42.45 &  0.17 &  Gold \\
 SN-04D3do & 0.610 &   42.98 &  0.17 &  Gold \\
 SN-04D3ez & 0.263 & 40.87 &  0.17 &  Gold \\
 SN-04D3fk & 0.358 & 41.66 &  0.17 &  Gold \\
 SN-04D3fq & 0.730 & 43.47 &  0.18 &  Gold \\
 SN-04D3gt & 0.451 & 42.22 &  0.17 &  Silver \\
 SN-04D3gx & 0.910 & 44.44  & 0.38 &  Silver \\
 SN-04D3hn & 0.552 & 42.65  & 0.17 &  Gold \\
 SN-04D3is & 0.710 & 43.36  & 0.18  & Silver \\
 SN-04D3ki & 0.930 & 44.61 &  0.46 &  Silver \\
 SN-04D3kr & 0.337 & 41.44 &  0.17  &      Gold \\
 SN-04D3ks & 0.752 & 43.35 &  0.19 &  Silver \\
 SN-04D3lp & 0.983 & 44.13 & 0.52 &  Silver \\
 SN-04D3lu  &  0.822  &  43.73 &  0.27 &  Gold \\
 SN-04D3ml  &  0.950  &  44.14 &  0.31 &  Gold \\
 SN-04D3nc  &  0.817  &  43.84 &  0.30 &       Silver \\
 SN-04D3nh  &  0.340  &  41.51 &  0.17 &  Gold \\
 SN-04D3nr  &  0.960  &    43.81 &    0.28 &  Silver \\
 SN-04D3ny  &  0.810  &  43.88 &  0.34 &  Silver \\
 SN-04D3oe  &  0.756  &  43.64 &  0.17 &  Gold \\
 SN-04D4an  &  0.613  &  43.15 &    0.18 &  Gold \\
 SN-04D4bk  &  0.840  &  43.66 &  0.25 &  Silver \\
 SN-04D4bq  &  0.550  &   42.67 &    0.17 &  Gold \\
 SN-04D4dm  &  0.811  &  44.13 &  0.31 &  Gold \\
 SN-04D4dw  &  0.961  &  44.18 &    0.33 &  Gold \\
\enddata
\end{deluxetable}

\begin{thebibliography}{99}
\bibitem[\protect\citeauthoryear{Amanullah et al.}{2010}]{Amanullah} Amanullah, R. et al., 2010, Astrophys. J., 716, 712.
\bibitem[\protect\citeauthoryear{Armendariz - Picon et al.}{2001}]{Armen} Armendariz - Picon, C. et al., 2001, Phys. Rev. D , 63, 103510.
\bibitem[\protect\citeauthoryear{Ashtekar}{2007}]{Ashtekar2007}Ashtekar, A., 2007, Nuovo Cim. B, 122, 135.
\bibitem[\protect\citeauthoryear{Ashtekar et al.}{2003}]{Ashtekar2003}Ashtekar, A. et al., 2003, Adv. Theor. Math. Phys., 7, 233.
\bibitem[\protect\citeauthoryear{Ashtekar et al.}{2004}]{Ashtekar}Ashtekar, A. et al., 2004, Class. Quantum. Grav., 21, R53.
\bibitem[\protect\citeauthoryear{Ashtekar et al.}{2006}]{Ashtekar2006}Ashtekar, A. et al., 2006, Class. Quantum. Grav., 23, 391.
\bibitem[\protect\citeauthoryear{Ashtekar et al.}{2008}]{Ashtekar2008}Ashtekar, A. et al., 2008, Phys. Rev. D, 77, 024046.
\bibitem[\protect\citeauthoryear{Ashtekar et al.}{2011}]{Ashtekar2011}Ashtekar, A. et al., 2011, Class. Quantum. Grav., 28, 213001.
\bibitem[\protect\citeauthoryear{Astier et al.}{2006}]{Astier} Astier, P. et al., 2006, Astron. Astrophys., 447, 31.
\bibitem[\protect\citeauthoryear{Bachall et al.}{1999}]{Bachall}Bachall, N. A. et al, 1999, Science 284, 1481.
\bibitem[\protect\citeauthoryear{Balart et al.}{2007}]{Balart}Balart, L. et al., 2007, Eur. Phys. J. C, 51, 185.
\bibitem[\protect\citeauthoryear{Bamba et al.}{2013}]{Bamba}Bamba, K. et al., 2013,  arXiv:1211.2968v2.
\bibitem[\protect\citeauthoryear{Barris et al.}{2004}]{Barris}Barris, B. J. et al., 2004, Astrophys. J., 602, 571.
\bibitem[\protect\citeauthoryear{Bennet et al.}{2000}]{Bennet}Bennet, C. et al, 2000, Phys. Rev. Lett. 85, 2236.
\bibitem[\protect\citeauthoryear{Bojowald}{2001}]{Bojowald2001}Bojowald, M., 2001, Phys. Rev. Lett., 86, 5227.
\bibitem[\protect\citeauthoryear{Bojowald}{2002}]{Bojowald2002}Bojowald, M., 2002, Phys. Rev. Lett., 89, 261301.
\bibitem[\protect\citeauthoryear{Bojowald}{2005}]{Bojowald2005}Bojowald, M., 2005, liv. Rev. Rel., 8, 11.
\bibitem[\protect\citeauthoryear{Bojowald}{2008}]{Bojowald2008}Bojowald, M., 2008, liv. Rev. Rel., 11, 4.
\bibitem[\protect\citeauthoryear{Bond et al.}{1997}]{Bond}Bond, J. R. et al, 1997, Mon. Not. Roy. Astron. Soc., 291, L33.
\bibitem[\protect\citeauthoryear{Bouhmadi-López et al.}{2010}]{Bouhmadi}Bouhmadi-López, M. and Chimento, L. P., 2010, Phys.Rev. D, 82, 103506.
\bibitem[\protect\citeauthoryear{Briddle et al.}{2003}]{Briddle}Briddle, S. et al, 2003, Science 299, 1532.
\bibitem[\protect\citeauthoryear{Cailleteau et al.}{2008}]{Cailleteau et al.}Cailleteau, T. et al., 2008, Phys. Rev. Lett 101, 251302.
\bibitem[\protect\citeauthoryear{Caldwell}{2002}]{Caldwell}Caldwell, R. R., 2002, Phys. Lett. B, 545, 23.
\bibitem[\protect\citeauthoryear{Caldwell et al.}{2003}]{Caldwell2003}Caldwell, R. R. et al., 2003, Phys. Rev. Lett., 91, 071301.
\bibitem[\protect\citeauthoryear{Chakraborty et al.}{2007}]{Writ}Chakraborty, W., Chakraborty, S. and Debnath, U., 2007, Grav. Cosmol, 13, 293.
\bibitem[\protect\citeauthoryear{Chakraborty et al.}{2012}]{Chak}Chakraborty, S., Debnath, U. and Ranjit, C., 2012, Eur. Phys. J. C, 72, 2101.
\bibitem[\protect\citeauthoryear{Chen et al.}{2008}]{Chen}Chen, S., Wang, B. and Jing, J., 2008, Phys. Rev. D, 78, 123503.
\bibitem[\protect\citeauthoryear{Choudhury et al.}{2007}]{Paddy1}Choudhury, T. R. and Padmanabhan, T., 2007, Astron. Astrophys., 429, 807.
\bibitem[\protect\citeauthoryear{Chowdhury et al.}{2013}]{Ratul}Chowdhury, R. et al., 2013, Int. J. Theor. Phys., 52,489.
\bibitem[\protect\citeauthoryear{Corichi et al.}{2009}]{Corichi}Corichi, A. et al., 2009, Phys. Rev. D, 80, 044024.
\bibitem[\protect\citeauthoryear{Dao-Jun et al.}{2005}]{Jun}Dao-Jun, L. and Xin-Zhou, L., 2005, Chin. Phys. Lett., 22, 1600.
\bibitem[\protect\citeauthoryear{Debnath et al.}{2004}]{Debnath}Debnath, U., Banerjee, A. and Chakraborty, S., 2004, Class. Quantum Grav., 21, 5609.
\bibitem[\protect\citeauthoryear{del Campo et al.}{2009}]{del}del Campo, S., Herrera, R., Toloza, A., 2009, Phys. Rev. D, 79, 083507.
\bibitem[\protect\citeauthoryear{Efstathiou et al.}{1999}]{Efstathiou}Efstathiou, G. and Bond, J. R., 1999, Mon. Not. R. Astro. Soc. 304, 75.
\bibitem[\protect\citeauthoryear{Eisenstein et al.}{2005}]{Eisenstein}Eisenstein, D. J. et al[SDSS Collaboration], 2005, Astrophys. J. 633, 560.
\bibitem[\protect\citeauthoryear{Farajollahi et al.}{2011}]{Farajollahi}Farajollahi, H., Ravanpak, A., 2011, Phys. Rev. D, 84, 084017.
\bibitem[\protect\citeauthoryear{Fu et al.}{2008}]{Fu} Fu, X., Yu, H. and Wu, P., 2008, Phys. Rev. D 78, 063001.
\bibitem[\protect\citeauthoryear{Ghose et al.}{2012}]{Paul3} Ghose, S., Thakur, P. and Paul, B. C., 2012, Mon. Not. R. Astron. Soc., 421, 20.
\bibitem[\protect\citeauthoryear{Gonzalez-Diaz}{2003}]{Gon} Gonz$\acute{a}$lez-Diaz, P. F., 2003, Phys. Rev. D, 68, 021303(R).
\bibitem[\protect\citeauthoryear{Gorini et al.}{2003}]{Gorini}Gorini, V.,  Kamenshchik, A. and Moschella, U., 2003, Phys. Rev. D, 67, 063509.
\bibitem[\protect\citeauthoryear{Jain et al.}{2003}]{Jain}Jain, B. and Taylor, A., 2003, Phys. Rev. Lett. 91, 141302.
\bibitem[\protect\citeauthoryear{Jamil et al.}{2011}]{jamil}Jamil, M. and Debnath, U., 2011, Astrophys Space Sci., 333, 3.
\bibitem[\protect\citeauthoryear{Kamenshchik et al.}{2001}]{Kamenshchik}Kamenshchik, A. et al., 2001, Phys. Lett. B, 511, 265.
\bibitem[\protect\citeauthoryear{Komatsu et al.}{2011}]{Komatsu}Komatsu, E. et al, 2011, Astrophys. J. Suppl., 192, 18.
\bibitem[\protect\citeauthoryear{Kowalaski et al.}{2008}]{Kowalaski}Kowalaski et al, 2008, Astrophys. J., 686, 749.
\bibitem[\protect\citeauthoryear{Lamon et al.}{2010}]{Lamon}Lamon, R. et al., 2010, Phys. Rev. D, 81, 024026.
\bibitem[\protect\citeauthoryear{Lu et al.}{2008}]{Lu}Lu, J. et al., 2008, Phys. Lett. B, 662, 87.
\bibitem[\protect\citeauthoryear{Malquarti et al.}{2003}]{Malquarti}Malquarti, M., et al., 2003, Phys. Rev. D, 68, 023512.
\bibitem[\protect\citeauthoryear{Marcus}{1990}]{Marcus}Marcus, N., 1990, Gen. Rel. Grav., 22, 873.
\bibitem[\protect\citeauthoryear{Martin et al.}{2008}]{Martin}Martin, J. and Yamaguchi, M., 2008, Phys. Rev. D, 77, 123508.
\bibitem[\protect\citeauthoryear{Miller et al.}{1999}]{Miller}Miller, D. et al, 1999, Astrophys. J. 524, L1.
\bibitem[\protect\citeauthoryear{Morris}{2012}]{Morris}Morris, J. R., 2012, Gen. Rel. Grav., 44, 437.
\bibitem[\protect\citeauthoryear{Naskar et al.}{2007}]{Naskar}Naskar, T. et al., 2007, Phys. Rev. D, 76, 063514.
\bibitem[\protect\citeauthoryear{Nessaeris et al.}{2007}]{Nessaeris}Nessaeris, S. and Perivolaropoulos, L., 2007, JCAP, 0701, 018.
\bibitem[\protect\citeauthoryear{Nojiri, et al.}{2005}]{Nojiri2005}Nojiri, S. et al., 2005, Phys. Rev. D, 71, 063004.
\bibitem[\protect\citeauthoryear{Nojiri, et al.}{2011}]{Nojiri2011}Nojiri, S. et al., 2011, Phys. Rept., 505, 59.
\bibitem[\protect\citeauthoryear{Padmanabhan}{2003}]{Paddy}Padmanabhan, T., 2003, Phys. Rept., 380, 235.
\bibitem[\protect\citeauthoryear{Padmanabhan et al.}{2003}]{Paddy2}Padmanabhan, T. and Choudhury, T. R., 2003, Mon. Not. R. Astron. Soc., 344, 823.
\bibitem[\protect\citeauthoryear{Paul et al.}{2010}]{Paul1} Paul, B. C., Thakur, P. and Ghose, S., 2010, Mon. Not. Roy. Astron. Soc., 407, 415.
\bibitem[\protect\citeauthoryear{Paul et al.}{2011}]{Paul2} Paul, B. C., Ghose, S. and Thakur, P., 2011, Mon. Not. R. Astron. Soc., 413, 686.
\bibitem[\protect\citeauthoryear{Peebles et al.}{1988}]{Peebles}Peebles, P. J. E. and Ratra, B., 1988, Astrophys. J. Lett., 325, L17.
\bibitem[\protect\citeauthoryear{Perlmutter et al.}{1998}]{Perlmutter}Perlmutter, S. J. et al, 1998, Nature 391, 51.
\bibitem[\protect\citeauthoryear{Perlmutter et al.}{1999}]{Perlmutter1}Perlmutter, S. J. et al[SNCP Collsboration], 1999, Astrophys. J., 517, 565.
\bibitem[\protect\citeauthoryear{Riess et al.}{1998}]{Riess}Riess, A. G. et al.[Supernova Search Team Collaboration], 1998, Astron. J. 116, 1009.
\bibitem[\protect\citeauthoryear{Riess et al.}{2004}]{Riess1}Riess, A. G. et al., 2004, Astrophys. J. 607, 665.
\bibitem[\protect\citeauthoryear{Riess et al.}{2007}]{Riess2}Riess, A. G. et al., 2007, Astrophys. J., 659, 98.
\bibitem[\protect\citeauthoryear{Rovelli}{1998}]{Rovelli}Rovelli, C, 1998, Liv. Rev. Rel., 1, 1.
\bibitem[\protect\citeauthoryear{Rudra et al.}{2012a}]{Rudra2012a}Rudra, P. et al., 2012, Astrophys.Space Sci. 339,53.
\bibitem[\protect\citeauthoryear{Rudra et al.}{2012b}]{Rudra2012b}Rudra, P. et al., 2012, Astrophys.Space Sci. 342,557.
\bibitem[\protect\citeauthoryear{Sadjadi}{2013}]{Sadjadi}Sadjadi, H. M., 2013, Eur. Phys. J. C, 73, 2571.
\bibitem[\protect\citeauthoryear{Sahni et al.}{2000}]{Sahni}Sahni, V. and Starobinsky, A. A., 2000, Int. J. Mod. Phys. D, 9, 373.
\bibitem[\protect\citeauthoryear{Samart et al.}{2007}]{Samart}Samart, D. et al., 2007, Phys. Rev. D, 76, 043514.
\bibitem[\protect\citeauthoryear{Sami et al.}{2006}]{Sami}Sami, M. et al., 2006, Phys. Rev. D, 74, 043514.
\bibitem[\protect\citeauthoryear{Sen}{2002}]{Sen}Sen, A., 2002, J. High Energy Phys. 0204, 48.
\bibitem[\protect\citeauthoryear{Singh}{2009}]{Singh}Singh, P., 2009, Class. Quant. Grav., 26, 125005.
\bibitem[\protect\citeauthoryear{Singh et al.}{2011}]{Singh et al.}Singh, P. et al., 2011, Phys. Rev. D, 83, 064027.
\bibitem[\protect\citeauthoryear{Spalinski}{2007}]{Spalinski}Spalinski, M., 2007, J. Cosmol. Astropart. Phys., 05, 017.
\bibitem[\protect\citeauthoryear{Spergel et al.}{2003}]{Spergel}Spergel, D. N. et al, 2003, Astrophys. J. Suppl. 148, 175.
\bibitem[\protect\citeauthoryear{Spergel et al.}{2007}]{Spergel1}Spergel, D. N. et al, 2007, Astrophys. J. Suppl. 170, 377.
\bibitem[\protect\citeauthoryear{Stern et al.}{2010}]{Stern}Stern, D. et al, 2010, JCAP, 1002, 008.
\bibitem[\protect\citeauthoryear{Tedmark et al.}{2004}]{Tedmark}Tedmark, M. et al, 2004, Phys. Rev. D 69, 103501.
\bibitem[\protect\citeauthoryear{Thakur et al.}{2009}]{Paul} Thakur, P., Ghose, S. and Paul, B. C., 2009, Mon. Not. R. Astron. Soc., 397, 1935.
\bibitem[\protect\citeauthoryear{Tonry et al.}{2003}]{Tonry}Tonry, J. L. et al., 2003, Astrophys. J., 594, 1.
\bibitem[\protect\citeauthoryear{Wei et al.}{2005}]{Wei}Wei, H., et al., 2005, Class. Quantum. Grav., 22, 3189.
\bibitem[\protect\citeauthoryear{Weinberg}{1989}]{Weinberg}Weinberg, S., 1989, Rev. Mod. Phys., 61, 1.
\bibitem[\protect\citeauthoryear{Wu et al.}{2007}]{Wu1}Wu, P. and Yu, H., 2007, Phys. Lett. B, 644, 16.
\bibitem[\protect\citeauthoryear{Wu et al.}{2008}]{Wu}Wu, P. and Zhang, S. N., 2008, JCAP, 06, 007.
\end{thebibliography}
\end{document}